\pdfoutput=1
\documentclass{aa}
\usepackage{graphicx}
\usepackage{txfonts}
\usepackage{placeins}
\usepackage{xspace}
\usepackage[colorlinks,allcolors=blue]{hyperref}


\newcommand{\Teff}{\ensuremath{T_{\rm eff}}\xspace}
\newcommand{\logg}{\ensuremath{\log g}\xspace}
\newcommand{\vsini}{\ensuremath{v\sin i}\xspace}
\newcommand{\vbroadcol}{\texttt{vbroad}\xspace}
\newcommand{\vbroad}{\texttt{vbroad}\xspace}
\newcommand{\Grvs}{\ensuremath{G_{\rm RVS}}\xspace}
\newcommand{\kms}{\ensuremath{\mathrm{km\,s^{-1}}}\xspace}
\newcommand{\dex}{\ensuremath{\mathrm{dex}}\xspace}

\newcommand{\ruwe}{\texttt{ruwe}\xspace}

\newcommand{\vbroaderror}{\texttt{vbroad\_error}\xspace}

\providecommand{\corrauth}[1]{\email{#1}}
\begin{document}
\nolinenumbers\renewcommand{\linenumbers}{}
\title{Empirical rotation velocity distributions from Gaia DR3 across the Kraft break}

\author{S.\ Dainese\inst{1}\corrauth{dainese@phys.au.dk}
  \and S.\ H.\ Albrecht\inst{1}\email{albrecht@phys.au.dk}
  \and M.\ L.\ Marcussen\inst{1}\email{marcus@phys.au.dk}
  \and E.\ Knudstrup\inst{1}\email{emil@phys.au.dk}}

\institute{Department of Physics and Astronomy, Aarhus University, Ny Munkegade 120, 8000 Aarhus C, Denmark}

\date{Submitted to A\&A on 30 November 2026}

\titlerunning{Rotation velocity distributions from Gaia DR3 across the Kraft break}
\authorrunning{S. Dainese et al.}

\abstract{We provide rotation velocity distributions for over two hundred thousand single main-sequence and subgiant stars from Gaia Data Release 3 (DR3), covering effective temperatures $6000$--$8000$~K and surface gravities $\logg = 3.5$--$4.5$. We measure how the Kraft break shifts with surface gravity. For over three million stars, Gaia DR3 gives the width of the spectral lines (\vbroadcol{}), which depends on the rotation velocity projected along the line of sight, \vsini{}. We calibrate \vbroadcol{} against \vsini{} from high-resolution spectra. We then bin the stars by effective temperature and surface gravity, in cells of $100~\mathrm{K} \times 0.1~\dex$, or $50~\mathrm{K} \times 0.05~\dex$ where there are many stars, and fit the rotation velocity distribution in each cell with a two-parameter Weibull function. We find that rotation rises steeply with temperature but also depends on surface gravity. Below about 6800~K, stars of a given temperature rotate slowest near $\logg \approx 4.0$ and faster toward both lower and higher surface gravity. The Kraft break begins at $\Teff = 6500 \pm 31$~K. In the full sample, rotation is halfway between the slow and fast levels at $6935 \pm 58$~K. The onset of the break depends on surface gravity and is hottest at $\logg = 4.0$--$4.1$. The fitted distributions can be used as rotation priors for planning observations and for population studies.}

\keywords{stars: rotation -- surveys -- methods: statistical}

\maketitle

\section{Introduction}\label{sec:intro}
Stars form from contracting clouds of gas. As the gas contracts, it keeps its angular momentum, so young stars rotate fast. Over their lifetime, some stars slow down again. For these stars, rotation and color together give an age, the basis of gyrochronology \citep{Skumanich1972, Barnes2003}. Rotation also matters for exoplanet studies. Fast rotators have broad spectral lines that limit radial-velocity precision \citep{Bouchy2001}, and the projected rotation velocity, \vsini{}, is a key input when measuring the spin--orbit angles of planet hosts and in binary star systems \citep{Albrecht2022}.

The projected rotation velocity is measured from the width of the spectral lines. The spectrum is integrated over the visible stellar disk, which includes both the approaching and the receding side, and the resulting Doppler shifts widen the spectral lines. The broadening is set by \vsini{}, where $v$ is the rotation velocity and $i$ is the inclination of the spin axis to the line of sight.

Main-sequence stars rotate at anything from below $1~\kms$ to several hundred \kms{}, depending on their mass and age. The Kraft break \citep{Kraft1967} divides main-sequence rotation into two regimes. Above about $6500$~K, rotation rises steeply with temperature \citep[e.g.,][]{Gray1984, Royer2007, Beyer2024}. Stars cooler than the break have deep convective envelopes that drive strong dynamos \citep{Brun2017}. The magnetic fields of cool stars couple to the ionized stellar wind, and the wind steadily drains their angular momentum \citep{Schatzman1962, Weber1967, Kawaler1988, Pinsonneault1989}. Hotter, more massive stars have thin outer convection zones or none, and in most cases no large-scale magnetic fields. Without dynamo-driven winds, they hardly slow down and remain rapidly rotating throughout their main-sequence lives \citep{Mestel1968}. \citet{Beyer2024} located the break in a volume-limited sample of nearby F stars, cleaned of young stars, evolved stars, and candidate multiples. Their result rests on a few hundred stars.

Mapping how rotation is distributed across the Hertzsprung--Russell diagram needs large, homogeneous samples. The classical compilations combined data from many instruments and covered limited ranges in temperature \citep[e.g.,][]{Abt1995, Glebocki2005}. Multiobject surveys now provide \vsini{} or line-broadening measurements for far larger samples \citep[e.g.,][]{Majewski2017, Zuo2024, Buder2025}, but each within its own footprint and target selection. The European Space Agency's Gaia mission \citep{GaiaPrusti2016} provides such a sample, from one instrument and one pipeline. Its third data release \citep[DR3;][]{GaiaVallenari2023} measures the width of the spectral lines of 3.5 million stars \citep{Fremat2023}. Rotation is not the only cause of line broadening, so the measured width does not map one-to-one onto \vsini{} (Sect.~\ref{sec:calibration}). \citet{Sussholz2026} used the measured widths directly and located the Kraft break by eye. We build on their work by calibrating the widths against \vsini{}, by accounting for the stars too slow for Gaia to publish a value, and by fitting the shape of the transition.

With this calibration, we infer the distribution of rotation velocities of single stars in cells of \Teff{} and \logg{}. From the same sample, we locate the Kraft break and measure the width of the transition. The grid gives an empirical prior on the rotation of a single star in each cell. Such priors help plan radial-velocity and Rossiter--McLaughlin observations. They are also useful in statistical obliquity studies. For example, \citet{Marcussen2024} measured the spin--orbit alignment of 2727 Gaia astrometric binaries by comparing their line broadening with their orbital inclinations. An aligned star has a \vsini{} near zero if its orbit is seen face-on, and shows its full rotation velocity if the orbit is seen edge-on. Such a comparison benefits from knowing the rotation velocities of similar single stars, which this paper provides.

We describe how Gaia measures line broadening (Sect.~\ref{sec:instrument-pipeline}), select single main-sequence and subgiant stars (Sect.~\ref{sec:gaia-sample}), and calibrate Gaia's line broadening against \vsini{} from high-resolution spectra (Sect.~\ref{sec:calibration}). We then model the distribution of rotation velocities in cells of \Teff{} and \logg{} (Sect.~\ref{sec:modeling-velocity-distributions}), present the resulting grid (Sect.~\ref{sec:results}), and locate the Kraft break (Sect.~\ref{sec:kraft}).

\section{Gaia's \texorpdfstring{\vbroadcol{}}{vbroad} parameter}\label{sec:instrument-pipeline}
Before we calibrate \vbroadcol{}, we describe how Gaia measures it and which stars it leaves out. Gaia's Radial Velocity Spectrometer (RVS) records the 846--870~nm window around the \ion{Ca}{ii} triplet at a median resolving power of 11\,500 \citep{Cropper2018, Katz2023}. For slowly rotating stars, most of the line width comes from the instrument, and rotation adds only a small part. Gaia can therefore measure slow rotation only in spectra with a high signal-to-noise ratio. The relative uncertainty of \vbroadcol{} is below 20\% for $\Grvs < 9$ and about 60\% at $\Grvs = 11$. Above $7000$~K, broad Paschen lines blend with the \ion{Ca}{ii} triplet and leave few features that respond to rotation \citep{Fremat2023}.

For each epoch, the Gaia DR3 pipeline broadens a template spectrum with a series of trial broadenings and interpolates between them to find the one that best matches the observed spectrum. The published \vbroadcol{} is the median of these per-epoch values over at least six epochs, and \vbroaderror{} is their standard deviation. Gaia DR3 gives no \vbroadcol{} value for stars whose broadening is at or below $5~\kms$ \citep{Fremat2023}.

\begin{figure}
\centering
\includegraphics{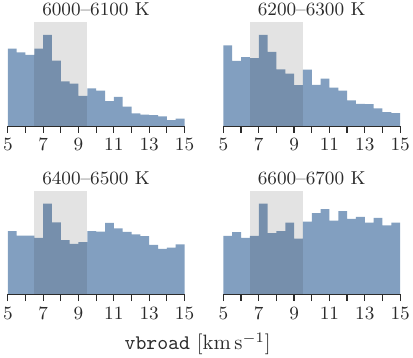}
\caption{Pile-up of \vbroadcol{} near the publication floor. Each panel shows one 100~K cell at $\logg = 4.0$--$4.1$. Bins are 0.5~\kms{} wide, and each panel has its own vertical scale. The shaded box marks $6.5$--$9.5~\kms$, the pile-up and the dip above it.}
\label{fig:pileup}
\end{figure}

Just above this floor, the \vbroadcol{} distribution shows a pile-up of measurements near $7$--$8~\kms$ (Fig.~\ref{fig:pileup}), with a deficit just above and a weaker excess near $10~\kms$. The excess over a smooth distribution amounts to about 9\% of all stars below $10~\kms$. The feature does not depend on \Grvs, metallicity, Galactic position, or proper motion. It appears at all temperatures and stays at $7$--$8~\kms$ while the distribution under it turns from falling to rising with temperature. We did not model this feature, and it limits how well we fit the slowest stars (Sect.~\ref{sec:grid-fitting}).

\section{Our sample}\label{sec:gaia-sample}
\begin{figure}
\centering
\includegraphics{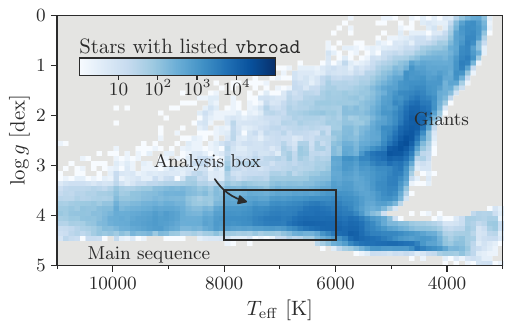}
    \caption{Kiel diagram of the 1\,950\,625 Gaia DR3 stars with a published \vbroadcol{} and GSP-Phot \Teff{} and \logg{}, 55\% of the catalog. The black box is our analysis box.}
  \label{fig:gaia-all}
\end{figure}

\begin{figure*}
\centering
\includegraphics{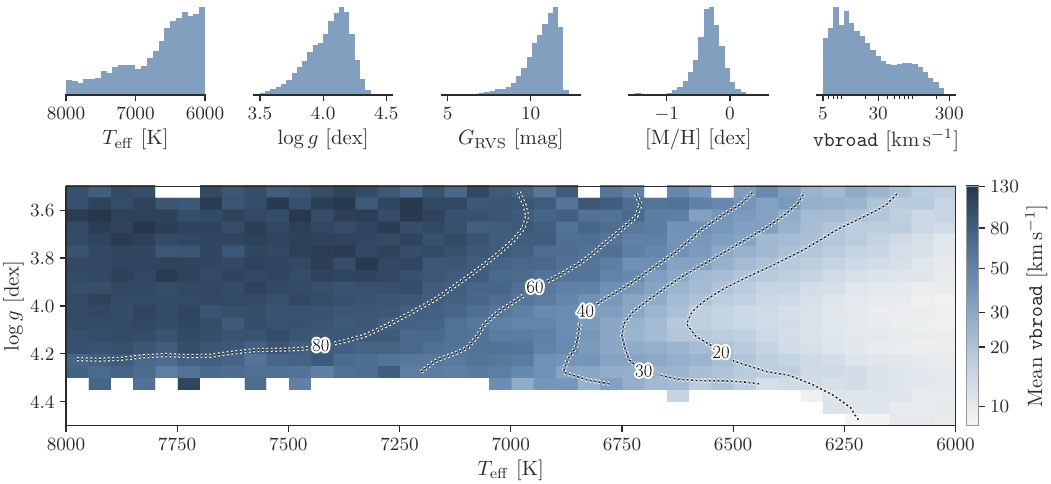}
\caption{Selected sample (Sect.~\ref{sec:gaia-sample}). The top row shows histograms of temperature, surface gravity, \Grvs, GSP-Phot metallicity, and \vbroadcol. The bottom panel shows the mean \vbroadcol{} in bins of $50~\mathrm{K} \times 0.05~\dex$, with smoothed contours of constant mean \vbroadcol{}.}
\label{fig:sample-intro}
\end{figure*}

We want single main-sequence and subgiant stars with reliable \vbroadcol{} on both sides of the Kraft break. Figure~\ref{fig:gaia-all} shows the Gaia \vbroadcol{} catalog in a Kiel diagram. It contains main-sequence stars over a wide range of temperatures, and cooler, lower-gravity subgiants and giants. We used the \Teff{} and \logg{} from the Gaia DR3 General Stellar Parametrizer from Photometry (GSP-Phot), which derives them from the low-resolution spectra of Gaia's blue and red photometers (BP/RP) \citep{Andrae2023}. We applied several filters that cut the sample from 3.5 million to 200\,000 (Table~\ref{tab:cascade}). We required each star to have a published \Grvs{} magnitude, a \Teff{} uncertainty below $100$~K, and a \vbroadcol{} between 5 and $275~\kms$. The upper limit of $275~\kms$ is below the range where the relation between \vbroadcol{} and \vsini{} becomes nonlinear \citep[$300$--$400~\kms$;][]{Fremat2023}. 

Next, we removed binaries, as a companion can change the stellar rotation speed. In a close binary, tides drive the rotation toward the orbital period, which for cool stars is often shorter than the rotation period of a comparable single star. A bright companion also adds its light, which shifts the star's color and brightness and with them the GSP-Phot \Teff{} and \logg{}. The companion's lines also blend with those of the primary, which broadens \vbroadcol{} \citep{Hadad2025}. Gaia publishes no \vbroadcol{} for stars whose spectra its pipeline flags as double-lined \citep{Katz2023, Fremat2023}. However, this flag could miss single-lined binaries and binaries whose components have similar radial velocities and line widths, and these still bias \vbroadcol{}. We excluded targets in the Gaia DR3 non-single-star catalogs \citep{GaiaArenou2023} and in the wide-binary catalog of \citet{ElBadry2021}, as listed by \citet{Hwang2022}, as well as stars with a renormalized unit weight error (\ruwe) above 1.2 \citep{Lindegren2021}, whose astrometry does not fit a single star. We also removed targets whose radial velocity varies between epochs,\footnote{We flagged a star when both \texttt{rv\_renormalised\_gof} $> 4$ and \texttt{rv\_chisq\_pvalue} $< 0.01$, following \citet{Katz2023}.} which catches spectroscopic binaries that the astrometry misses.

Broad lines make radial velocities less precise, so we checked that this screen does not simply remove fast rotators. Among stars at $5500$--$8000$~K with \vbroadcol{} between 50 and $100~\kms$, 96\% of those at $5500$--$6000$~K have variable radial velocities, but only 6\% of those at $7400$--$8000$~K, where fast rotation is common. The screen therefore mostly removes binaries rather than fast rotators.

Finally, we restricted the sample to the analysis box, $6000 < \Teff < 8000$~K and $3.5 < \logg < 4.5$ (Figs.~\ref{fig:gaia-all} and \ref{fig:sample-intro}). Below $6000$~K many stars are expected to rotate too slowly for Gaia to publish a \vbroadcol{}.
Above $8000$~K only a few stars are in our sample, and their \vbroadcol{} values are less precise. Our surface gravity limits keep main-sequence stars and subgiants and remove giants. For the fit of the Kraft break (Sect.~\ref{sec:kraft}), we extended the lower limit to 5500~K.

We applied no metallicity cut and did not test whether rotation depends on metallicity, because the GSP-Phot metallicities are systematically too low \citep{Andrae2023} and the metallicities from the RVS spectra (GSP-Spec) are derived assuming no rotation \citep{RecioBlanco2023}. Kinematically the sample belongs to the thin disk, with 99\% of the stars having peculiar velocities below the $85~\kms$ thin-disk limit of \citet{Fuhrmann2004} and \citet{Nissen2004}. After these cuts, 208\,387 stars remain in the analysis box. The 202\,290 of them that fall in cells with at least 250 stars (Sect.~\ref{sec:modeling-velocity-distributions}) form the sample of the grid (Table~\ref{tab:cascade}).

\begin{table}
\caption{Sample selection.}
\label{tab:cascade}
\centering
\begin{tabular}{lr}
\hline\hline
\noalign{\smallskip}
Cut & Stars \\
\hline
\noalign{\smallskip}
Published \vbroadcol{} in DR3 & 3\,524\,677 \\
Quality, completeness, and binary filters & 1\,296\,214 \\
Radial-velocity variability screen & 1\,179\,930 \\
Analysis box ($6000$--$8000$~K, $\logg = 3.5$--$4.5$) & 208\,387 \\
Cells with $\ge$250 stars (final, 292 cells) & 202\,290 \\
\hline
\end{tabular}
\end{table}

\section{From \texorpdfstring{\vsini{}}{v sin i} to \texorpdfstring{\vbroadcol{}}{vbroad}}
\label{sec:calibration}
Gaia broadens its template with a rotational profile whose only free parameter is the velocity (Sect.~\ref{sec:instrument-pipeline}), so any other extra width ends up in the \vbroadcol{} value. This includes macroturbulence, as well as errors from template mismatch and from the assumed instrumental line spread \citep{Fremat2023}. We therefore calibrated \vbroadcol{} against \vsini{} measured from high-resolution spectra. The calibration predicts the \vbroadcol{} that Gaia would report for a star with a given \vsini{}.

We built the calibration sample from the 516 stars of \citet{Marcussen2024}, observed with the FIbre-fed Echelle Spectrograph \citep[FIES;][]{Telting2014} at the Nordic Optical Telescope \citep[NOT;][]{Djupvik2010}. \citet{Marcussen2024} measured \vsini{} with iSpec \citep{BlancoCuaresma2014}. We excluded 170 of these stars by the single-star criteria of Sect.~\ref{sec:gaia-sample}. Of these, 107 belong to wide binaries \citep{ElBadry2021}, 4 are in the Gaia DR3 non-single-star catalogs, 42 more have $\ruwe > 1.2$, and 17 more have variable radial velocities. This leaves 346 stars. Above about $30~\kms$, the iSpec values fall increasingly below those from a broadening function \citep{Rucinski1992,Kaluzny2006}, so we measured \vsini{} again with a broadening function for 50 of the 55 stars above $30~\kms$.\footnote{We used synthetic templates \citep{Meszaros2024} and a linear limb-darkening law, with the coefficient fitted under a Gaussian prior of $0.68 \pm 0.10$. We also fitted a Gaussian macroturbulence, with a prior of width $2~\kms$ centered on the relation of \citet{Doyle2014}, which we extrapolate above $6400$~K.} The \citet{Marcussen2024} sample reaches only $158~\kms$. We added two fast rotators with high \vbroadcol{} and archival FIES spectra, HD\,19655 and HD\,158788, measured with the same broadening function. With them, the calibration has 348 stars and reaches $\vsini = 220~\kms$. We treated \vsini{} below $2~\kms$ as unresolved. For the 53 stars below this limit, we assumed that every \vsini{} between 0 and $2~\kms$ is equally likely.\footnote{In Eq.~(\ref{eq:tobit}), the term for each star below $2~\kms$ is averaged over 0 to $2~\kms$ before taking the logarithm.} Of the 348 calibration stars, 269 have a Gaia \vbroadcol{} value.

\begin{figure}
\centering
\includegraphics{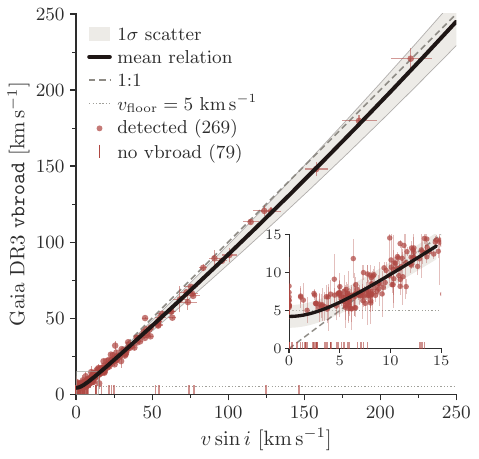}
\caption{Gaia DR3 \vbroadcol{} against \vsini{} for the 348 NOT stars. Filled circles are the 269 stars with a \vbroadcol{} value. Vertical error bars show the quoted Gaia \vbroadcol{} uncertainties, for display only, and horizontal error bars the \vsini{} uncertainties of the stars measured with a broadening function. Ticks on the horizontal axis mark the 79 stars without a \vbroadcol{} value, 38 of which are below $2~\kms$ and treated as unresolved. The dotted line is the $5~\kms$ publication floor. The solid line is the mean relation of Eq.~\eqref{eq:model-b} and the shaded band its $1\sigma$ scatter from Eq.~\eqref{eq:sigma-b}. The dashed line is the one-to-one relation. The inset is a zoom of the gray box, $0$--$15~\kms$ on both axes.}
\label{fig:calibration}
\end{figure}

Figure~\ref{fig:calibration} shows \vbroadcol{} versus \vsini{} for the 269 calibration stars with \vbroadcol{}. Rotation and the other sources of line broadening add roughly in quadrature, so we modeled the mean \vbroadcol{}, $\mu$, of stars with a given \vsini{} as
\begin{equation}\label{eq:model-b}
\mu(\vsini) = \sqrt{\mu_0^2 +(a\,\vsini)^2} + c\left(\frac{\vsini}{100~\kms}\right)^{2}.
\end{equation}
Here $\mu_0$ is the mean \vbroadcol{} of a nonrotating star, $a$ is the slope of the relation, and $c$ lets the relation steepen at high \vsini{}. Above about $20~\kms$, $\mu_0$ hardly matters and the relation is close to $\mu = a\,\vsini + c\,(\vsini/100~\kms)^2$. The last term adds $4~\kms$ at $\vsini = 100~\kms$ and $20~\kms$ at $220~\kms$, the fastest calibration star. The fitted $c$ differs from zero by $2\sigma$.

We assumed Gaussian scatter around the mean, with variance
\begin{equation}\label{eq:sigma-b}
\sigma^{2}(\vsini) \;=\; \sigma_0^{\,2} \;+\; (\sigma_1\,\vsini)^{2},
\end{equation}
where $\sigma_0$ is the scatter for a nonrotating star and $\sigma_1$ is the fraction of \vsini{} by which the scatter grows for faster rotators. This scatter comes from real differences in line broadening between stars of similar \vsini{}, such as differences in macroturbulence, and from the measurement uncertainties of FIES and Gaia.

We fitted the parameters to all 348 stars. For each star, Eqs.~\eqref{eq:model-b} and \eqref{eq:sigma-b} give the mean $\mu_i$ and width $\sigma_i$ of its predicted \vbroadcol{} distribution, and Gaia either reports a value (269 stars) or does not (79 stars).
The 79 stars without a \vbroadcol{} still carry information, since a missing value is expected for a slow star and surprising for a fast one. Leaving these stars out would keep only the slow stars that scattered above the publication floor, $v_{\rm floor} = 5~\kms$, and bias the relation toward high \vbroadcol{} at low \vsini{}.
For each of the 79 stars, we computed the probability $\Phi_i$ that its \vbroadcol{} lies below the floor, assuming that the Gaussian scatter we see above the floor continues below it. However, not every star without a value is too slow. Gaia also leaves out some stars for reasons unrelated to the publication floor \citep[see][for details]{Fremat2023}. We described these omissions with a single probability $f$, the same for all stars, that a star is
left out, regardless of its \vsini{}. The probability that a star's \vbroadcol{} lies above the floor but Gaia left it out is then $f\,(1-\Phi_i)$, and the probability of having no value is the sum of the two, $\Phi_i + f\,(1-\Phi_i)$.

A star with a published \vbroadcol{} enters with the probability density of that value, times $1-f$, the probability that it was not left out. The log-likelihood is
\begin{align}\label{eq:tobit}
\log \mathcal{L}(\theta)
={}& \sum_{i \in \mathrm{det}}
      \log\!\left[(1-f)\,\mathcal{N}\!\left(\vbroad_i \,\middle|\, \mu_i,\,\sigma_i\right)\right] \nonumber \\
   &+ \sum_{i \in \mathrm{none}}
      \log\!\left[\Phi_i + f\,(1-\Phi_i)\right],
\end{align}
where $\Phi_i = \Phi\big((v_{\rm floor} - \mu_i)/\sigma_i\big)$, $\mathcal{N}$ is the Gaussian probability density, $\Phi$ is the cumulative distribution function of the standard normal distribution, and $\theta = \{a, \mu_0, c, \sigma_0, \sigma_1, f\}$. The first sum runs over the 269 stars with a \vbroadcol{} value and the second over the 79 without. For stars faster than about $10~\kms$, $\Phi_i$ is almost zero, so a fast star that lacks a \vbroadcol{} was left out for another reason. Such stars set $f$. The fit attributes about 58 of the 79 missing values to the publication floor and the rest to other reasons.

We sampled the six parameters with \texttt{emcee} \citep{ForemanMackey2013}, using flat priors.\footnote{The flat priors keep $a$, $\mu_0$, $\sigma_0$, and $\sigma_1$ positive, because each enters the model only as a square and a negative value would give the same relation. The prior on $c$ allows either sign, and the prior on $f$ is uniform between 0 and 1. We ran 64 walkers for 6000 steps, more than 50 autocorrelation times, and discarded the first 1500 as burn-in.} The posterior medians and $1\sigma$ intervals are
\begin{equation}\label{eq:calibration-posterior}
\begin{aligned}
a        &= 0.875 \pm 0.012, & \sigma_{0} &= 1.49 \pm 0.11~\kms,\\
\mu_{0}  &= 4.23^{+0.19}_{-0.20}~\kms, & \sigma_{1} &= 0.063 \pm 0.008,\\
c        &= 4.2 \pm 2.1~\kms, & f &= 0.076^{+0.019}_{-0.017}.
\end{aligned}
\end{equation}
The residuals of the relation show no trend with temperature or with GSP-Phot $[\mathrm{M/H}]$, so we included neither term.

Of the 269 stars with both \vsini{} and \vbroadcol{}, 260 lie between $6000$ and $7000$~K. All are brighter than $\Grvs = 12$ and almost all brighter than $\Grvs = 10$, against a median of $10.9$ in our sample. Only 15 rotate faster than $70~\kms$. Their \vbroadcol{} lies on average $6.2 \pm 1.1~\kms$ below their remeasured \vsini{}, in line with \citet[][Sect.~4.3]{Fremat2023}, who find no overestimate of \vbroadcol{} for fast rotators. The fastest of them has $\vsini = 220$~\kms{}, while the grid uses \vbroadcol{} up to $275$~\kms{}. For hotter cells, fainter stars, and faster rotators, the grid relies on the calibration beyond the range of its stars. The relation has no magnitude term, because the calibration stars cannot constrain one, so Eq.~\eqref{eq:sigma-b} likely underestimates the scatter for the fainter stars, whose Gaia uncertainties are larger (Sect.~\ref{sec:instrument-pipeline}).

\section{Modeling rotation velocities} \label{sec:modeling-velocity-distributions}

\begin{figure}
\centering
\includegraphics{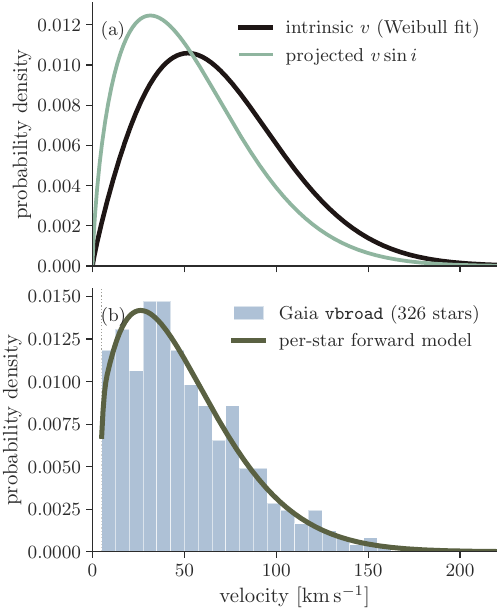}
\caption{Modeling chain for the $6800 \le \Teff < 6850$~K and $3.90 \le \logg < 3.95$ cell. (a) Fitted rotation velocity distribution ($k = 1.88 \pm 0.11$, $\lambda = 77.9 \pm 2.9~\kms$) and its projection to \vsini{} for isotropic inclinations (Eq.~\eqref{eq:vsini-projection}). (b) Resulting \vbroadcol{} density of Eq.~\eqref{eq:cell-likelihood}, including the calibration, its scatter, and the $5$--$275~\kms$ window (curve), compared with the cell's 326 stars (histogram). The dotted line marks the $5~\kms$ floor.}
\label{fig:simulation-method}
\end{figure}

With the calibration in place, we can construct a grid of rotation velocity distributions across \Teff{} and \logg{} that can serve, for example, as a lookup table of empirical rotation priors (Sect.~\ref{sec:rotation-velocity-grid}). We divided the analysis box into small cells and fitted the distribution of rotation velocities $v$ in each cell. This distribution cannot be built star by star, because the inclinations are unknown, \vbroadcol{} scatters about the calibration relation (Sect.~\ref{sec:calibration}), and the sample contains no stars with \vbroadcol{} outside 5--$275~\kms$ (Sects.~\ref{sec:instrument-pipeline} and \ref{sec:gaia-sample}). We therefore analyzed all stars of a cell together. We assumed that the rotation velocities $v$ in a cell follow a single distribution with two free parameters. We computed the distribution of \vbroadcol{} it would produce, taking into account random inclinations, the calibration, and the window (Sect.~\ref{sec:per-star-likelihood}). We then adjusted the two parameters until the predicted distribution best matched the measured one (Sect.~\ref{sec:grid-fitting}). Figure~\ref{fig:simulation-method} shows these steps for one cell.

We used cells of $100~\mathrm{K} \times 0.1~\dex$ and kept cells with at least 250 stars, enough to constrain the two parameters of the velocity distribution (Sect.~\ref{sec:per-star-likelihood}). Where each quarter of a cell holds at least 250 stars, we split the cell into four cells of $50~\mathrm{K} \times 0.05~\dex$. The resulting 292 cells hold 97\% of the stars in the region selected in Sect.~\ref{sec:gaia-sample}.

\begin{figure*}[!t]
\centering
\includegraphics{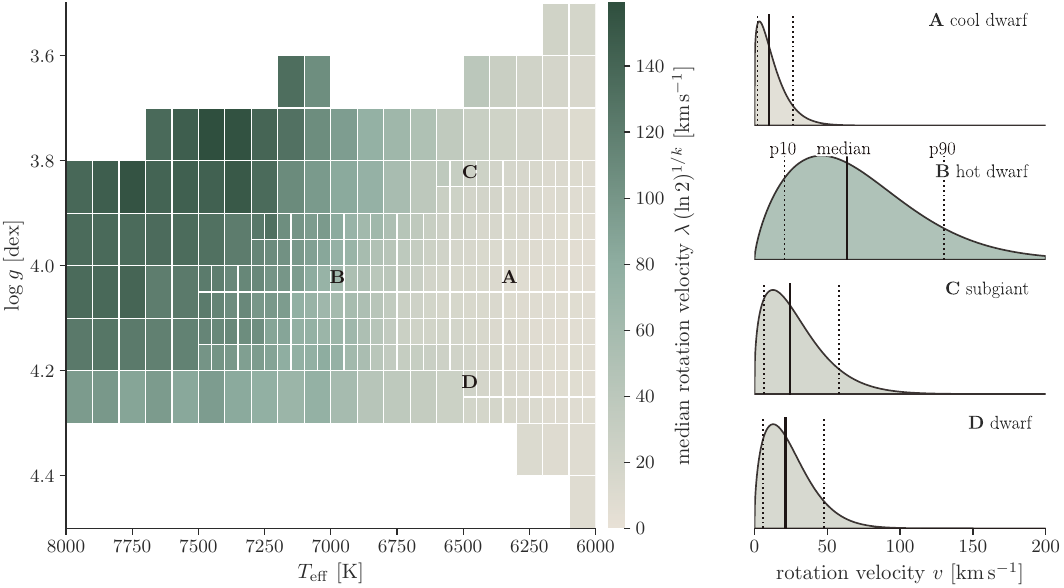}
\caption{Median rotation velocity of the best-fit Weibull distribution in each cell (left), blank where a cell has fewer than 250 stars. The right panels show the fitted distributions of the cells marked A to D, with the median (solid) and the 10th and 90th percentiles (dotted). A and B share surface gravity, and C and D share temperature.}
\label{fig:velocity-grid}
\end{figure*}

\subsection{A model for each cell}
\label{sec:per-star-likelihood}

We described the rotation velocities in each cell with a Weibull distribution,
\begin{equation}
p(v) = \frac{k}{\lambda} \left( \frac{v}{\lambda} \right)^{k-1} \exp \!\left[ -\left( \frac{v}{\lambda} \right)^k \right], \quad v \geq 0, \label{eq:weibull}
\end{equation}
where the scale $\lambda$ sets the typical velocity and the shape $k$ sets how narrow and how skewed the distribution is. For $k = 1$ the Weibull distribution is exponential, for $k = 2$ it is the Rayleigh distribution, and it becomes more symmetric as $k$ grows. These two parameters describe both the narrow, skewed distributions of the cool stars, where most stars rotate slowly with a tail of faster rotators, and the broad, more symmetric distributions of the hot stars. 

For isotropic spin axes \citep{Chandrasekhar1950}, the \vsini{} of stars whose rotation velocities follow Eq.~\eqref{eq:weibull} has the density
\begin{equation} \label{eq:vsini-projection}
p(\vsini \mid k, \lambda) = \int_{0}^{\pi/2} p\left(\frac{\vsini}{\sin i}\right) \mathrm{d}i ,
\end{equation}
where $i$ is the inclination of the spin axis. The calibration of Sect.~\ref{sec:calibration} then turns each \vsini{} into a Gaussian distribution of \vbroadcol{} with mean $\mu$ and variance $\sigma^2$ (Eqs.~\eqref{eq:model-b} and \eqref{eq:sigma-b}). 

Unlike in the calibration, where the 79 stars without a value are known, we do not know how many stars of a cell fall outside the 5--$275~\kms$ window. We therefore fit only the shape of the distribution inside the window, dividing each star's predicted density by the fraction of the model that falls inside it,
\begin{align}
p(\vbroad_j \mid k, \lambda) &= Z^{-1}\!\!\int\!\!\mathcal{N}\big(\vbroad_j \mid \mu(\vsini), \nonumber \\
&\quad \sigma(\vsini)\big)\, p(\vsini \mid k, \lambda)\, \mathrm{d}(\vsini), \nonumber \\
Z(k, \lambda) &= \!\int\! P\big(5 \le \vbroad \le 275 \mid \vsini\big) \nonumber \\
&\quad \times p(\vsini \mid k, \lambda)\, \mathrm{d}(\vsini),
\label{eq:cell-likelihood}
\end{align}
where $Z$ is the probability that a star drawn from the model lands inside the window, and $f_{\rm out} = 1 - Z$ is the fraction of the model distribution outside it, which we list for every cell. The log-likelihood of a cell is the sum of $\log p(\vbroad_j \mid k, \lambda)$ over its stars $j$, and we evaluated the integrals numerically.\footnote{We integrated over inclination with 40-node Gauss--Legendre quadrature and over \vsini{} on a 1000-point grid. The grid ran to twice the largest \vbroadcol{} in the cell divided by $a$, limited to 50--650~\kms.}

\subsection{Fitting and testing each cell} \label{sec:grid-fitting}
For every cell, we found the $k$ and $\lambda$ whose predicted \vbroadcol{} distribution best matches the observed one, by maximizing the log-likelihood of Sect.~\ref{sec:per-star-likelihood}. We took the $1\sigma$ uncertainties of $k$ and $\lambda$, and their correlation $\rho_{k\lambda}$, from the width of the likelihood peak. For the $2\sigma$ intervals and the posterior draws, we computed the posterior on a fine grid of $k$ and $\lambda$ around the best fit, with a flat prior on $k$ and $\lambda$.

To test the fit, we compared in each cell the observed and the predicted fraction of stars below each \vbroadcol{}. The largest difference between the two, the Kolmogorov--Smirnov distance $D_{\rm KS}$, is 0.043 in a typical cell, so the model is off by at most about 4\%. This misfit is small, but larger than noise alone would give. Artificial cells drawn from the best-fit models, with the same number of stars, have a typical $D_{\rm KS}$ of 0.026. In 172 of the 292 cells, the real misfit is larger than in 95\% of the artificial cells ($p < 0.05$). Appendix~\ref{app:weibull-appendix} lists $D_{\rm KS}$ and $p$ for every cell.

Part of the misfit comes from the pile-up near $7~\kms$ (Sect.~\ref{sec:instrument-pipeline}, Fig.~\ref{fig:pileup}), which we did not model. The pile-up mostly affects cells below 6800~K, where 53\% of stars have a \vbroadcol{} below 12~\kms{}, compared with 6\% in hotter cells. When we refit these cooler cells using only stars above 12~\kms{}, the misfit drops by a quarter, after allowing for the smaller number of stars, but stays larger than in the hotter cells.\footnote{We compared $D_{\rm KS}\sqrt{N}$, which for a correct model does not depend on the number of stars $N$.} The model mainly misses the detailed distribution of the slowest stars between 5 and 12~\kms{}.

\section{Rotation velocity grid} \label{sec:results}
Figure~\ref{fig:velocity-grid} shows the median rotation velocity, $\lambda (\ln 2)^{1/k}$, of the best-fit distribution in each cell. Table~\ref{tab:grid-excerpt} gives $(k, \lambda)$ for the first three cells, and Appendix~\ref{app:weibull-appendix} lists all 292. Rotation rises sharply with temperature. At $\logg \ge 4.0$, the median velocity goes from about $7~\kms$ at $6000$--$6100$~K to $98~\kms$ above $7000$~K, and the distribution broadens and shifts to higher velocities from the cool dwarf A to the hot dwarf B. At fixed temperature, subgiants rotate faster than dwarfs, as seen for the subgiant C and the dwarf D. Below $6500$~K, for example, the median velocity is $17~\kms$ at $\logg \le 3.9$ and $8~\kms$ at $\logg = 4.0$--$4.1$. The faster rotation of the cool subgiants may reflect evolution. Stars hotter than the onset of the Kraft break were hardly braked. As they evolve they expand and cool, so the low-gravity cool cells may mix spun-down stars with fast rotators from above the break. The same pattern bends the contours in Fig.~\ref{fig:sample-intro} toward cooler temperatures at low gravity. The trend with gravity is not monotonic, however. Below about $6800$~K, rotation is slowest near $\logg \approx 4.0$ and rises again toward higher gravity. At $6400$--$6500$~K, the median velocity is $14~\kms$ at $\logg = 4.0$--$4.1$ and $19~\kms$ at $\logg = 4.2$--$4.3$.

Across the grid, the shape parameter spans $k = 1.0$--$2.6$ and the scale spans $\lambda = 7$--$194~\kms$. In cells at $\Teff \le 6300$~K, the distributions are narrow and moderately skewed, with a median $k$ of 1.3. In cells at $\Teff \ge 6700$~K, they are broad, with a median $k$ of 1.83, between the cool-cell value and the Rayleigh case ($k = 2$).
\begin{table}
\caption{Velocity distribution parameters for the first three cells of the grid.}
\label{tab:grid-excerpt}
\centering
\begin{tabular}{llcc}
\hline\hline
\noalign{\smallskip}
$T_{\rm eff}$ & $\log g$ & $k$ & $\lambda$ \\
(K) & (dex) & & (km\,s$^{-1}$) \\
\hline
\noalign{\smallskip}
6000--6100 & 3.50--3.60 & $1.51 \pm 0.10$ & $22.59 \pm 1.14$ \\
6000--6100 & 3.60--3.70 & $1.43 \pm 0.07$ & $18.31 \pm 0.69$ \\
6000--6100 & 3.70--3.80 & $1.36 \pm 0.06$ & $13.02 \pm 0.44$ \\
\ldots & \ldots & \ldots & \ldots \\
\hline
\end{tabular}
\tablefoot{The parameters $k$ and $\lambda$ are the shape and scale of the fitted Weibull distribution of rotation velocities (Eq.~\eqref{eq:weibull}), with $1\sigma$ uncertainties. The full table is given in Appendix~\ref{app:weibull-appendix} (Table~\ref{tab:grid_full}), which lists all 292 cells and also gives the number of stars and the fit statistics of each cell.}
\end{table}

\subsection{Comparison with measured \texorpdfstring{\vsini{}}{v sin i}} \label{sec:louden}
To test the grid against \vsini{} measured without Gaia, we projected our fitted distributions to \vsini{} and compared them with \citet{Louden2021}. They selected a control sample of Kepler-field stars without regard to rotation, with \vsini{} from high-resolution spectra and spectroscopic binaries removed. They find a median \vsini{} of $4.5 \pm 0.4~\kms$ at $6000$--$6150$~K, where our cells at $\logg \ge 4.0$ give $5.20 \pm 0.05~\kms$, $1.7\sigma$ above their value. At $6150$--$6350$~K they find 7.0~\kms{}, and we find $7.23 \pm 0.05~\kms$. The grid thus agrees with this independent spectroscopy on the cool side, close to the 5~\kms{} floor.

\subsection{How to use the rotation velocity grid} \label{sec:rotation-velocity-grid}
The grid of the 292 fitted distributions is a lookup table for empirical rotation priors. For a given star, take the cell that contains its \Teff{} and \logg{}. The cell's $(k, \lambda)$ in Eq.~\eqref{eq:weibull} give the distribution of rotation velocities of the stars in that cell, and Eq.~\eqref{eq:vsini-projection} turns this distribution into a \vsini{} distribution for random inclinations. The grid is on GitHub.\footnote{\url{https://github.com/SilkeDainese/gaia-dr3-rotation-grid}} The cells are in GSP-Phot \Teff{} and \logg{}, which in our sample lie a median of 96~K below GSP-Spec temperatures. A spectroscopic temperature should therefore be shifted to the GSP-Phot scale before looking up a cell. For each cell we also provide posterior draws of $(k, \lambda)$, which include the correlation between $k$ and $\lambda$ (Appendix~\ref{app:weibull-appendix}). The grid is specific to the Gaia DR3 sample as selected in Sect.~\ref{sec:gaia-sample} and extrapolates outside the calibration range of Sect.~\ref{sec:calibration}. The fitted distributions depend on magnitude. In a typical cell, $k$ and $\lambda$ are about 10\% lower in the fainter half, so the grid reflects the magnitude mix of our sample, not a volume-complete population.

\section{The Kraft break} \label{sec:kraft}
The grid shows that rotation rises steeply with \Teff{}, and that at fixed temperature it also depends on \logg{} (Sect.~\ref{sec:results}). This rise is the Kraft break. To give its position and width as numbers that can be compared with earlier work, we described the break with a single model across \Teff{}. We fit individual stars, not the cells of the grid. The cells are 100~K wide, wider than the precision we reach on the break, and a fit to stars lets each star enter with its own \Teff{} uncertainty. We first fit all stars together and then repeated the fit in slices of \logg{}. We fit \vsini{} directly, without converting to $v$. For random inclinations the mean \vsini{} is a fixed fraction, $\pi/4$, of the mean $v$, so the conversion would scale both levels by the same factor and leave the position and width of the break unchanged.

\subsection{Fitting the break} \label{sec:kraft-break}
\begin{figure}[!htb]
\centering
\includegraphics{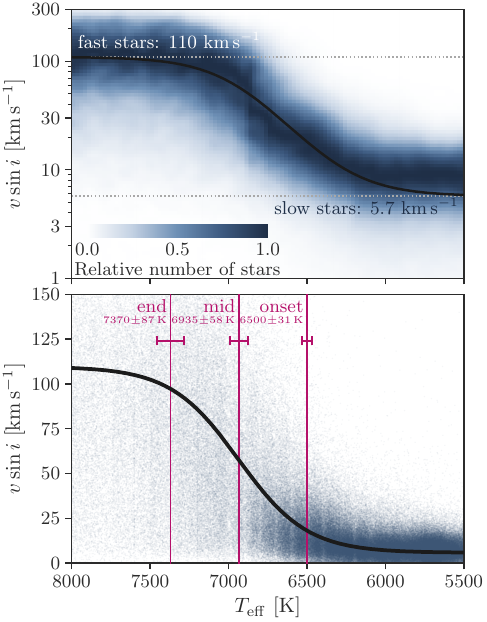}
\caption{Kraft break in the Gaia sample. Both panels show \vsini{} against \Teff{}. Each star is shown at one \vsini{} drawn from the range its \vbroadcol{} allows under the calibration (Sect.~\ref{sec:calibration}), so together the points trace the population, not the \vsini{} of single stars. The fit itself uses \vbroadcol{}. Top: density of stars. The color shows the number of stars per bin, relative to the most populated bin at each temperature. The black curve is the fitted model. Bottom: same stars as individual points. The magenta lines mark the onset, midpoint, and end of the break, with their uncertainties as horizontal bars. The model includes the stars below the $5~\kms$ publication floor, which are not plotted. At the cool end, the curve therefore lies below the mean of the plotted stars.}
\label{fig:kraft-break}
\end{figure}

Figure~\ref{fig:kraft-break} shows the Kraft break in our sample. Cool stars rotate slowly, and hot stars rotate fast. At each temperature, we described \vsini{} by a single Weibull distribution, whose mean and shape move from a constant level for the cool stars to a constant level for the hot stars along a sigmoid,
\begin{equation}
S(T_{\rm eff}) = \left[ 1 + \exp\!\left( -\frac{T_{\rm eff} - T_{\rm mid}}{w} \right) \right]^{-1}.
\end{equation}
The mean \vsini{} is then $(1-S)\,\bar{v}_{\rm cool} + S\,\bar{v}_{\rm hot}$, and the shape follows the same rule. Here $T_{\rm mid}$ is the temperature where the mean is halfway between the two levels, and $w$ sets the width of the transition. We defined the onset of the break as $T_{\rm onset} = T_{\rm mid} - 2w$, where the mean has moved 12\% of the way from the cool to the hot level. This is where rotation starts to rise. We defined the end as $T_{\rm end} = T_{\rm mid} + 2w$, where it has moved 88\% of the way. The fitted distribution also covers the stars below the 5~\kms{} floor, which Gaia does not publish. Each star enters with its \Teff{} uncertainty and the calibration of Sect.~\ref{sec:calibration}. The fitted mean \vsini{} is $5.7~\kms$ on the cool side and $110~\kms$ on the hot side.

We fit the 321\,185 stars that pass the cuts of Sect.~\ref{sec:gaia-sample} at $5500$--$8000$~K and $3.5 < \logg < 4.5$, by maximum likelihood. We extended the range below 6000~K so that the cool level is set by stars well below the break. The onset is at $6500 \pm 31$~K. The midpoint is at $6935 \pm 58$~K, and the transition ends at $7370 \pm 87$~K. The end lies above 7000~K and the midpoint just below it, where few calibration stars constrain the relation (Sect.~\ref{sec:calibration}). The onset lies inside the calibrated range. The fit itself gives the onset an uncertainty of only 2~K, and narrowing the fitted range to 6000--7500~K moves it by 3~K. Moving the calibration by its $1\sigma$ uncertainty moves the onset by 8~K. The largest term comes from the stars without a published \vbroadcol{}. Adding them to the fit moves the onset by $-29$~K. Added in quadrature, these give 31~K.

\begin{figure}[!t]
\centering
\includegraphics{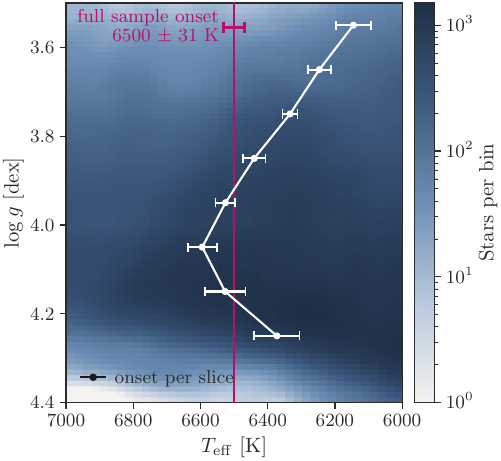}
\caption{The Kraft break as a function of surface gravity. The white circles show the onset fitted separately in each $0.1$~dex slice of \logg{}. The magenta line shows the Kraft break for the full sample (Fig.~\ref{fig:kraft-break}). Error bars show the uncertainties. The blue background color shows the number of stars per bin of $50~{\rm K} \times 0.05$~dex. Slices with too few hot stars to locate the break are not shown.}
\label{fig:break-per-logg}
\end{figure}

In slices of 0.1~dex in \logg{}, the onset depends on surface gravity (Fig.~\ref{fig:break-per-logg}). The onset is hottest, $6595 \pm 43$~K, at $\logg = 4.0$--$4.1$. At the same gravity, the grid shows the slowest rotation at fixed temperature (Sect.~\ref{sec:results}). Both results say that stars near $\logg = 4.0$ stay slow up to the hottest temperatures. Of our stars, 40\% lie at $\logg = 4.0$--$4.2$. These stars dominate the fit, so the onset of the full sample is hotter than that of most slices.

\subsection{Comparison with earlier work}\label{subsec:disk-kraft}
Table~\ref{tab:kraft-comparison} compares our onset, midpoint, and end (Sect.~\ref{sec:kraft-break}) with earlier work. Earlier studies typically give a single temperature or a narrow range near 6500~K. \citet{Kraft1967} found the drop in rotation near spectral type F5. \citet{Beyer2024} used a volume-limited sample within 33~pc and defined the Kraft break as the range in which the \vsini{} distribution is roughly uniform, between the distinct distributions of slow and fast rotators. They placed this range at $6450$--$6650$~K, with an uncertainty of 50~K on each edge. \citet{Sussholz2026} placed the break at about 6500~K by eye in Gaia \vbroadcol{}.

The onset is the only quantity with a counterpart in the earlier studies. It matches the value of \citet{Sussholz2026}, who used the same GSP-Phot temperature scale. It lies 50~K above the cool edge of the range of \citet{Beyer2024}, whose temperatures come from colors and spectroscopy. Temperatures from colors and spectroscopy differ from GSP-Phot by more than the 31~K uncertainty of our onset, so our onset and their edge cannot be compared at that precision.

The hot edge of \citet{Beyer2024} marks the end of their uniform range, not where the mean \vsini{} reaches the hot level, so it does not correspond to our end. In our fit the mean \vsini{} at 6700~K, which is above their hot edge, has moved only a quarter of the way to the fast level. Their 200~K range is therefore not a width of the transition as we define it, which is $T_{\rm end} - T_{\rm onset} = 4w = 870 \pm 60$~K.

The onset changes with surface gravity in our sample (Fig.~\ref{fig:break-per-logg}), so a comparison must account for the gravities in each sample. \citet{Beyer2024} kept only main-sequence stars and removed stars with $\logg < 3.75$ as evolved, while our sample reaches $\logg = 3.5$. We expect that our slice at $\logg = 4.2$--$4.3$ holds the least evolved stars, so it is the closest match to their sample. Its onset, $6373 \pm 67$~K, lies 77~K below the cool edge of their range, a difference of $0.9\sigma$. Mixing slices with different onsets may also widen the transition of the full sample. At low gravity, the cooler onset agrees with the picture of Sect.~\ref{sec:results}, in which stars that formed hotter than the break expand and cool as they evolve and keep their fast rotation. This picture does not explain the cooler onset at high gravity. How the onset changes with surface gravity may be set by the different evolutionary tracks that overlap in each slice. 

\begin{table}
\caption{Kraft-break determinations compared.}
\label{tab:kraft-comparison}
\centering
\setlength{\tabcolsep}{4pt}
\begin{tabular}{lrll}
\hline\hline
\noalign{\smallskip}
Study & $N$ & Definition & $T$ \\
 & & & (K) \\
\hline
\noalign{\smallskip}
\citet{Kraft1967}\tablefootmark{a} & \ldots & rotation drop & F5 \\
\citet{Beyer2024}\tablefootmark{b} & 278 & zone edges & 6450--6650 \\
\citet{Sussholz2026}\tablefootmark{c} & 350\,000 & by eye & $\sim$6500 \\
This work\tablefootmark{d} & 321\,185 & onset & $6500\pm31$ \\
 & & midpoint & $6935\pm58$ \\
 & & end & $7370\pm87$ \\
\hline
\noalign{\smallskip}
\multicolumn{4}{c}{This work, onset by $\logg$ slice} \\
\noalign{\smallskip}
\hline
\noalign{\smallskip}
$3.5$--$3.6$ & 4\,282 & & $6145\pm52$ \\
$3.6$--$3.7$ & 8\,272 & & $6246\pm34$ \\
$3.7$--$3.8$ & 15\,033 & & $6333\pm22$ \\
$3.8$--$3.9$ & 25\,600 & & $6440\pm34$ \\
$3.9$--$4.0$ & 38\,341 & & $6526\pm29$ \\
$4.0$--$4.1$ & 56\,799 & & $6595\pm43$ \\
$4.1$--$4.2$ & 70\,837 & & $6527\pm60$ \\
$4.2$--$4.3$ & 51\,815 & & $6373\pm67$ \\
\hline
\end{tabular}
\tablefoot{
\tablefoottext{a}{Given as a spectral type.}
\tablefoottext{b}{Volume-limited sample within 33~pc, $\logg \ge 3.75$, spectroscopic temperatures.}
\tablefoottext{c}{Gaia \vbroadcol{} with no correction for the 5~\kms{} floor, GSP-Phot temperatures.}
\tablefoottext{d}{Onset $= T_{\rm mid} - 2w$, where the mean \vsini{} has moved 12\% of the way from the slow to the fast level. End $= T_{\rm mid} + 2w$, where it has moved 88\% of the way. GSP-Phot temperatures.}
}
\end{table}

\section{Conclusion} \label{sec:conclusion}

We calibrated Gaia DR3 \vbroadcol{} against \vsini{} from high-resolution FIES spectra and used it to map the rotation of 202\,290 single main-sequence and subgiant stars. We divided the range $6000$--$8000$~K and $\logg = 3.5$--$4.5$ into small cells and fitted a Weibull distribution of rotation velocities in each. Rotation rises steeply with temperature, from median rotation velocities near 7~\kms{} in the cool main-sequence cells to 98~\kms{} above 7000~K. At fixed temperature below about 6800~K, rotation is slowest near $\logg \approx 4.0$ and faster toward both lower and higher surface gravity. The shape of the distribution changes as well. On the cool side, it is narrow, with most stars rotating slowly and a tail of faster rotators, and on the hot side it is broad.

The Kraft break begins at $6500 \pm 31$~K. Rotation is halfway between the slow and fast levels at $6935 \pm 58$~K, and the transition ends at $7370 \pm 87$~K and is in total $870 \pm 60$~K wide. The onset depends on surface gravity and is hottest, $6595 \pm 43$~K, at $\logg = 4.0$--$4.1$, and the mix of evolutionary states in our sample may widen the transition. The grid describes the Gaia DR3 sample after our cuts, and serves as a rotation prior for stars in the analysis box. 

\begin{acknowledgements}
S.D. and S.H.A. acknowledge support from the Independent Research Fund Denmark (Danmarks Frie Forskningsfond) through grant 5241-00072B.

This work has made use of data from the European Space Agency (ESA) mission Gaia (\url{https://www.cosmos.esa.int/gaia}), processed by the Gaia Data Processing and Analysis Consortium (DPAC; \url{https://www.cosmos.esa.int/web/gaia/dpac/consortium}). Funding for the DPAC has been provided by national institutions, in particular the institutions participating in the Gaia Multilateral Agreement.

Based on observations made with the Nordic Optical Telescope, owned in collaboration by the University of Turku and Aarhus University, and operated jointly by Aarhus University, the University of Turku and the University of Oslo, representing Denmark, Finland and Norway, the University of Iceland and Stockholm University at the Observatorio del Roque de los Muchachos, La Palma, Spain, of the Instituto de Astrofisica de Canarias. The FIES spectra used for the calibration were obtained under programs 60-803, 61-510, 66-503, 67-003, 67-261, and 68-009.

Facilities: Gaia \citep{GaiaPrusti2016}, NOT (FIES) \citep{Djupvik2010}. Software: emcee \citep{ForemanMackey2013}, iSpec \citep{BlancoCuaresma2014}.
\end{acknowledgements}

\bibliographystyle{aa}
\bibliography{references}

\begin{appendix}
\onecolumn
\section{The grid data product}\label{app:weibull-appendix}
Table~\ref{tab:grid_full} lists the full grid, one row per cell, with the cell bounds in \Teff and \logg, the number of stars $N$, the Weibull parameters $k$ and $\lambda$ with $1\sigma$ uncertainties, the posterior correlation $\rho(k,\lambda)$, the fraction $f_{\rm out}$ of the fitted distribution outside the $5$--$275~\kms$ window, the goodness-of-fit distance $D_{\rm KS}$, its $p$-value from simulated cells, and the percentage $p_{78}$ of the fitted model in the $7$--$8~\kms$ band around the pile-up, which can be compared with the observed stars to see how strongly the pile-up affects a cell. The same table is also available in machine-readable form, together with the $2\sigma$ intervals, posterior draws for direct resampling, and a status flag marking cells outside the range of the calibration stars (Sect.~\ref{sec:calibration}).
{\small
\begin{longtable}{llrccccccc}
\caption{Per-cell Weibull fits for all 292 grid cells.}\label{tab:grid_full}\\
\hline\hline
\noalign{\smallskip}
$T_{\rm eff}$ & $\log g$ & $N$ & $k$ & $\lambda$ & $\rho_{k\lambda}$ & $f_{\rm out}$ & $D_{\rm KS}$ & $p$ & $p_{78}$ \\
(K) & (dex) & & & (km\,s$^{-1}$) & & & & & (\%) \\
\hline
\noalign{\smallskip}
\endfirsthead
\caption{continued.}\\
\hline\hline
\noalign{\smallskip}
$T_{\rm eff}$ & $\log g$ & $N$ & $k$ & $\lambda$ & $\rho_{k\lambda}$ & $f_{\rm out}$ & $D_{\rm KS}$ & $p$ & $p_{78}$ \\
(K) & (dex) & & & (km\,s$^{-1}$) & & & & & (\%) \\
\hline
\noalign{\smallskip}
\endhead
\hline
\endfoot
6000--6100 & 3.50--3.60 & 309 & $1.51 \pm 0.10$ & $22.59 \pm 1.14$ & 0.51 & 0.10 & 0.038 & 0.280 & 6.6 \\
 & 3.60--3.70 & 683 & $1.43 \pm 0.07$ & $18.31 \pm 0.69$ & 0.58 & 0.14 & 0.026 & 0.450 & 8.3 \\
 & 3.70--3.80 & 1041 & $1.36 \pm 0.06$ & $13.02 \pm 0.44$ & 0.65 & 0.20 & 0.041 & $<0.010$ & 10.9 \\
6000--6050 & 3.80--3.85 & 345 & $1.70 \pm 0.14$ & $11.95 \pm 0.60$ & 0.66 & 0.19 & 0.030 & 0.680 & 12.4 \\
 & 3.85--3.90 & 423 & $1.56 \pm 0.11$ & $10.70 \pm 0.53$ & 0.68 & 0.23 & 0.042 & 0.050 & 13.2 \\
 & 3.90--3.95 & 505 & $1.20 \pm 0.07$ & $9.01 \pm 0.52$ & 0.69 & 0.30 & 0.081 & $<0.010$ & 13.1 \\
 & 3.95--4.00 & 662 & $1.29 \pm 0.08$ & $8.09 \pm 0.43$ & 0.76 & 0.32 & 0.055 & $<0.010$ & 14.2 \\
 & 4.00--4.05 & 808 & $1.25 \pm 0.06$ & $8.32 \pm 0.39$ & 0.74 & 0.31 & 0.059 & $<0.010$ & 13.9 \\
 & 4.05--4.10 & 1005 & $1.27 \pm 0.06$ & $8.45 \pm 0.35$ & 0.74 & 0.31 & 0.052 & $<0.010$ & 13.8 \\
 & 4.10--4.15 & 1172 & $1.21 \pm 0.05$ & $8.81 \pm 0.34$ & 0.73 & 0.30 & 0.071 & $<0.010$ & 13.3 \\
 & 4.15--4.20 & 1228 & $1.30 \pm 0.05$ & $8.73 \pm 0.32$ & 0.73 & 0.30 & 0.060 & $<0.010$ & 13.8 \\
 & 4.20--4.25 & 1305 & $1.12 \pm 0.04$ & $8.68 \pm 0.34$ & 0.73 & 0.31 & 0.067 & $<0.010$ & 12.9 \\
 & 4.25--4.30 & 1435 & $1.13 \pm 0.04$ & $9.46 \pm 0.35$ & 0.72 & 0.29 & 0.043 & $<0.010$ & 12.4 \\
6000--6100 & 4.30--4.40 & 2296 & $1.32 \pm 0.04$ & $11.00 \pm 0.27$ & 0.69 & 0.24 & 0.041 & $<0.010$ & 12.2 \\
 & 4.40--4.50 & 434 & $1.49 \pm 0.10$ & $11.11 \pm 0.56$ & 0.67 & 0.22 & 0.066 & $<0.010$ & 12.6 \\
6050--6100 & 3.80--3.85 & 314 & $1.49 \pm 0.11$ & $12.73 \pm 0.72$ & 0.64 & 0.20 & 0.042 & 0.220 & 11.4 \\
 & 3.85--3.90 & 388 & $1.15 \pm 0.07$ & $11.43 \pm 0.73$ & 0.65 & 0.25 & 0.063 & 0.010 & 11.3 \\
 & 3.90--3.95 & 437 & $1.46 \pm 0.10$ & $9.46 \pm 0.52$ & 0.72 & 0.27 & 0.053 & 0.010 & 13.9 \\
 & 3.95--4.00 & 570 & $1.24 \pm 0.08$ & $8.24 \pm 0.47$ & 0.75 & 0.32 & 0.054 & 0.010 & 13.8 \\
 & 4.00--4.05 & 714 & $0.98 \pm 0.04$ & $7.39 \pm 0.39$ & 0.64 & 0.36 & 0.093 & $<0.010$ & 12.9 \\
 & 4.05--4.10 & 877 & $1.41 \pm 0.07$ & $9.11 \pm 0.36$ & 0.71 & 0.28 & 0.062 & $<0.010$ & 14.0 \\
 & 4.10--4.15 & 956 & $1.20 \pm 0.05$ & $9.07 \pm 0.39$ & 0.72 & 0.30 & 0.047 & $<0.010$ & 13.1 \\
 & 4.15--4.20 & 996 & $1.31 \pm 0.06$ & $9.20 \pm 0.37$ & 0.73 & 0.28 & 0.042 & $<0.010$ & 13.5 \\
 & 4.20--4.25 & 1118 & $1.18 \pm 0.05$ & $9.50 \pm 0.38$ & 0.72 & 0.29 & 0.043 & $<0.010$ & 12.7 \\
 & 4.25--4.30 & 1172 & $1.32 \pm 0.05$ & $11.12 \pm 0.39$ & 0.69 & 0.24 & 0.035 & 0.010 & 12.1 \\
6100--6200 & 3.50--3.60 & 261 & $1.43 \pm 0.10$ & $28.31 \pm 1.60$ & 0.49 & 0.08 & 0.048 & 0.140 & 5.3 \\
 & 3.60--3.70 & 525 & $1.45 \pm 0.07$ & $23.89 \pm 0.94$ & 0.49 & 0.10 & 0.047 & $<0.010$ & 6.3 \\
 & 3.70--3.80 & 968 & $1.44 \pm 0.06$ & $18.26 \pm 0.56$ & 0.57 & 0.14 & 0.049 & $<0.010$ & 8.3 \\
6100--6150 & 3.80--3.85 & 357 & $1.56 \pm 0.11$ & $15.39 \pm 0.76$ & 0.61 & 0.15 & 0.038 & 0.280 & 9.8 \\
 & 3.85--3.90 & 484 & $1.65 \pm 0.11$ & $13.43 \pm 0.56$ & 0.63 & 0.17 & 0.032 & 0.290 & 11.1 \\
 & 3.90--3.95 & 591 & $1.17 \pm 0.06$ & $9.87 \pm 0.51$ & 0.66 & 0.28 & 0.086 & $<0.010$ & 12.4 \\
 & 3.95--4.00 & 722 & $1.23 \pm 0.06$ & $9.61 \pm 0.46$ & 0.71 & 0.28 & 0.064 & $<0.010$ & 12.8 \\
 & 4.00--4.05 & 861 & $1.13 \pm 0.05$ & $8.46 \pm 0.40$ & 0.71 & 0.32 & 0.081 & $<0.010$ & 13.1 \\
 & 4.05--4.10 & 993 & $1.42 \pm 0.06$ & $9.22 \pm 0.34$ & 0.70 & 0.28 & 0.064 & $<0.010$ & 13.9 \\
 & 4.10--4.15 & 1100 & $1.08 \pm 0.04$ & $8.69 \pm 0.35$ & 0.67 & 0.31 & 0.070 & $<0.010$ & 12.7 \\
 & 4.15--4.20 & 1226 & $1.23 \pm 0.05$ & $9.13 \pm 0.34$ & 0.71 & 0.29 & 0.062 & $<0.010$ & 13.1 \\
 & 4.20--4.25 & 1321 & $1.13 \pm 0.04$ & $10.22 \pm 0.39$ & 0.71 & 0.28 & 0.037 & $<0.010$ & 12.0 \\
 & 4.25--4.30 & 1163 & $1.30 \pm 0.05$ & $12.50 \pm 0.43$ & 0.67 & 0.22 & 0.041 & $<0.010$ & 11.1 \\
6100--6200 & 4.30--4.40 & 1134 & $1.32 \pm 0.05$ & $12.75 \pm 0.43$ & 0.66 & 0.21 & 0.048 & $<0.010$ & 11.0 \\
6150--6200 & 3.80--3.85 & 380 & $1.37 \pm 0.08$ & $15.88 \pm 0.83$ & 0.59 & 0.16 & 0.064 & $<0.010$ & 9.4 \\
 & 3.85--3.90 & 491 & $1.28 \pm 0.07$ & $12.78 \pm 0.65$ & 0.63 & 0.21 & 0.061 & $<0.010$ & 10.9 \\
 & 3.90--3.95 & 625 & $1.44 \pm 0.08$ & $11.85 \pm 0.50$ & 0.66 & 0.22 & 0.039 & 0.020 & 11.9 \\
 & 3.95--4.00 & 705 & $1.45 \pm 0.07$ & $10.79 \pm 0.43$ & 0.67 & 0.23 & 0.062 & $<0.010$ & 12.8 \\
 & 4.00--4.05 & 762 & $1.31 \pm 0.07$ & $9.51 \pm 0.43$ & 0.72 & 0.28 & 0.038 & 0.030 & 13.3 \\
 & 4.05--4.10 & 958 & $1.44 \pm 0.06$ & $10.43 \pm 0.37$ & 0.68 & 0.24 & 0.057 & $<0.010$ & 13.0 \\
 & 4.10--4.15 & 1052 & $1.18 \pm 0.04$ & $10.34 \pm 0.40$ & 0.66 & 0.27 & 0.083 & $<0.010$ & 12.1 \\
 & 4.15--4.20 & 1232 & $1.31 \pm 0.05$ & $9.73 \pm 0.33$ & 0.69 & 0.27 & 0.063 & $<0.010$ & 13.1 \\
 & 4.20--4.25 & 1359 & $1.13 \pm 0.04$ & $10.78 \pm 0.39$ & 0.69 & 0.27 & 0.072 & $<0.010$ & 11.6 \\
 & 4.25--4.30 & 1111 & $1.27 \pm 0.05$ & $12.80 \pm 0.45$ & 0.66 & 0.22 & 0.051 & $<0.010$ & 10.9 \\
6200--6300 & 3.60--3.70 & 449 & $1.50 \pm 0.08$ & $27.61 \pm 1.14$ & 0.48 & 0.08 & 0.039 & 0.110 & 5.3 \\
 & 3.70--3.80 & 1148 & $1.57 \pm 0.05$ & $24.03 \pm 0.60$ & 0.50 & 0.09 & 0.026 & 0.120 & 6.1 \\
6200--6250 & 3.80--3.85 & 421 & $1.58 \pm 0.10$ & $18.72 \pm 0.81$ & 0.56 & 0.12 & 0.023 & 0.900 & 8.0 \\
 & 3.85--3.90 & 551 & $1.64 \pm 0.09$ & $15.95 \pm 0.60$ & 0.59 & 0.14 & 0.050 & 0.010 & 9.5 \\
 & 3.90--3.95 & 715 & $1.29 \pm 0.06$ & $11.77 \pm 0.52$ & 0.67 & 0.23 & 0.039 & 0.020 & 11.6 \\
 & 3.95--4.00 & 718 & $1.40 \pm 0.07$ & $11.36 \pm 0.46$ & 0.67 & 0.23 & 0.051 & $<0.010$ & 12.2 \\
 & 4.00--4.05 & 784 & $1.08 \pm 0.04$ & $9.49 \pm 0.46$ & 0.67 & 0.30 & 0.065 & $<0.010$ & 12.1 \\
 & 4.05--4.10 & 908 & $1.39 \pm 0.06$ & $11.16 \pm 0.40$ & 0.66 & 0.23 & 0.061 & $<0.010$ & 12.3 \\
 & 4.10--4.15 & 1113 & $1.19 \pm 0.04$ & $10.33 \pm 0.37$ & 0.62 & 0.26 & 0.092 & $<0.010$ & 12.2 \\
 & 4.15--4.20 & 1451 & $1.46 \pm 0.05$ & $11.73 \pm 0.32$ & 0.64 & 0.21 & 0.074 & $<0.010$ & 12.1 \\
 & 4.20--4.25 & 1451 & $1.30 \pm 0.04$ & $12.70 \pm 0.38$ & 0.65 & 0.21 & 0.057 & $<0.010$ & 11.0 \\
 & 4.25--4.30 & 962 & $1.27 \pm 0.05$ & $15.41 \pm 0.56$ & 0.62 & 0.18 & 0.041 & 0.010 & 9.5 \\
6200--6300 & 4.30--4.40 & 540 & $1.38 \pm 0.08$ & $14.93 \pm 0.67$ & 0.62 & 0.18 & 0.049 & $<0.010$ & 9.9 \\
6250--6300 & 3.80--3.85 & 503 & $1.39 \pm 0.07$ & $20.10 \pm 0.87$ & 0.54 & 0.13 & 0.040 & 0.070 & 7.6 \\
 & 3.85--3.90 & 619 & $1.56 \pm 0.08$ & $17.76 \pm 0.65$ & 0.58 & 0.13 & 0.024 & 0.570 & 8.5 \\
 & 3.90--3.95 & 762 & $1.55 \pm 0.07$ & $15.82 \pm 0.53$ & 0.58 & 0.15 & 0.053 & $<0.010$ & 9.5 \\
 & 3.95--4.00 & 786 & $1.47 \pm 0.07$ & $12.78 \pm 0.46$ & 0.63 & 0.20 & 0.059 & $<0.010$ & 11.3 \\
 & 4.00--4.05 & 885 & $1.27 \pm 0.05$ & $11.23 \pm 0.44$ & 0.64 & 0.24 & 0.082 & $<0.010$ & 11.9 \\
 & 4.05--4.10 & 926 & $1.30 \pm 0.05$ & $11.56 \pm 0.44$ & 0.65 & 0.23 & 0.048 & $<0.010$ & 11.7 \\
 & 4.10--4.15 & 1105 & $1.56 \pm 0.06$ & $13.08 \pm 0.37$ & 0.61 & 0.18 & 0.072 & $<0.010$ & 11.3 \\
 & 4.15--4.20 & 1487 & $1.63 \pm 0.05$ & $13.70 \pm 0.32$ & 0.58 & 0.17 & 0.070 & $<0.010$ & 10.9 \\
 & 4.20--4.25 & 1454 & $1.37 \pm 0.04$ & $15.27 \pm 0.41$ & 0.59 & 0.17 & 0.079 & $<0.010$ & 9.7 \\
 & 4.25--4.30 & 749 & $1.44 \pm 0.06$ & $17.73 \pm 0.63$ & 0.57 & 0.14 & 0.047 & $<0.010$ & 8.5 \\
6300--6400 & 3.60--3.70 & 400 & $1.56 \pm 0.08$ & $35.87 \pm 1.46$ & 0.43 & 0.05 & 0.055 & 0.010 & 3.8 \\
 & 3.70--3.80 & 1329 & $1.50 \pm 0.05$ & $29.00 \pm 0.69$ & 0.47 & 0.07 & 0.029 & 0.030 & 5.1 \\
6300--6350 & 3.80--3.85 & 569 & $1.43 \pm 0.07$ & $22.48 \pm 0.88$ & 0.52 & 0.11 & 0.040 & 0.040 & 6.8 \\
 & 3.85--3.90 & 668 & $1.47 \pm 0.06$ & $20.53 \pm 0.72$ & 0.52 & 0.11 & 0.046 & $<0.010$ & 7.4 \\
 & 3.90--3.95 & 809 & $1.49 \pm 0.07$ & $16.70 \pm 0.55$ & 0.58 & 0.14 & 0.049 & 0.010 & 9.0 \\
 & 3.95--4.00 & 869 & $1.37 \pm 0.05$ & $14.89 \pm 0.51$ & 0.57 & 0.17 & 0.066 & $<0.010$ & 9.9 \\
 & 4.00--4.05 & 934 & $1.23 \pm 0.05$ & $13.30 \pm 0.50$ & 0.62 & 0.21 & 0.077 & $<0.010$ & 10.5 \\
 & 4.05--4.10 & 975 & $1.32 \pm 0.05$ & $13.40 \pm 0.47$ & 0.62 & 0.20 & 0.067 & $<0.010$ & 10.6 \\
 & 4.10--4.15 & 1166 & $1.44 \pm 0.05$ & $14.05 \pm 0.41$ & 0.60 & 0.18 & 0.056 & $<0.010$ & 10.4 \\
 & 4.15--4.20 & 1513 & $1.49 \pm 0.05$ & $15.29 \pm 0.38$ & 0.59 & 0.16 & 0.055 & $<0.010$ & 9.8 \\
 & 4.20--4.25 & 1422 & $1.50 \pm 0.05$ & $17.99 \pm 0.44$ & 0.56 & 0.13 & 0.055 & $<0.010$ & 8.4 \\
 & 4.25--4.30 & 654 & $1.53 \pm 0.07$ & $18.53 \pm 0.65$ & 0.55 & 0.12 & 0.065 & $<0.010$ & 8.1 \\
6350--6400 & 3.80--3.85 & 489 & $1.60 \pm 0.08$ & $25.34 \pm 0.94$ & 0.47 & 0.08 & 0.038 & 0.130 & 5.7 \\
 & 3.85--3.90 & 641 & $1.67 \pm 0.08$ & $23.12 \pm 0.74$ & 0.50 & 0.08 & 0.028 & 0.380 & 6.3 \\
 & 3.90--3.95 & 707 & $1.37 \pm 0.06$ & $19.48 \pm 0.72$ & 0.55 & 0.13 & 0.051 & $<0.010$ & 7.8 \\
 & 3.95--4.00 & 773 & $1.18 \pm 0.04$ & $15.92 \pm 0.66$ & 0.58 & 0.19 & 0.077 & $<0.010$ & 9.1 \\
 & 4.00--4.05 & 877 & $1.48 \pm 0.06$ & $15.71 \pm 0.50$ & 0.57 & 0.16 & 0.067 & $<0.010$ & 9.5 \\
 & 4.05--4.10 & 1022 & $1.44 \pm 0.05$ & $16.10 \pm 0.49$ & 0.57 & 0.16 & 0.058 & $<0.010$ & 9.3 \\
 & 4.10--4.15 & 1107 & $1.43 \pm 0.05$ & $15.77 \pm 0.47$ & 0.59 & 0.16 & 0.061 & $<0.010$ & 9.5 \\
 & 4.15--4.20 & 1335 & $1.71 \pm 0.06$ & $17.74 \pm 0.40$ & 0.54 & 0.12 & 0.045 & $<0.010$ & 8.5 \\
 & 4.20--4.25 & 1209 & $1.36 \pm 0.04$ & $19.84 \pm 0.56$ & 0.54 & 0.13 & 0.067 & $<0.010$ & 7.7 \\
 & 4.25--4.30 & 447 & $1.56 \pm 0.09$ & $22.57 \pm 0.92$ & 0.51 & 0.10 & 0.048 & 0.010 & 6.6 \\
6400--6500 & 3.60--3.70 & 387 & $1.58 \pm 0.08$ & $51.55 \pm 2.07$ & 0.39 & 0.03 & 0.036 & 0.280 & 2.4 \\
 & 3.70--3.80 & 1083 & $1.43 \pm 0.05$ & $37.26 \pm 1.00$ & 0.44 & 0.06 & 0.045 & $<0.010$ & 3.9 \\
6400--6450 & 3.80--3.85 & 488 & $1.65 \pm 0.08$ & $29.75 \pm 1.06$ & 0.45 & 0.06 & 0.038 & 0.110 & 4.7 \\
 & 3.85--3.90 & 576 & $1.56 \pm 0.07$ & $25.55 \pm 0.90$ & 0.48 & 0.08 & 0.028 & 0.460 & 5.7 \\
 & 3.90--3.95 & 733 & $1.43 \pm 0.06$ & $19.75 \pm 0.70$ & 0.55 & 0.12 & 0.042 & 0.010 & 7.7 \\
 & 3.95--4.00 & 768 & $1.27 \pm 0.05$ & $17.84 \pm 0.69$ & 0.56 & 0.16 & 0.075 & $<0.010$ & 8.4 \\
 & 4.00--4.05 & 920 & $1.43 \pm 0.05$ & $17.31 \pm 0.54$ & 0.52 & 0.14 & 0.089 & $<0.010$ & 8.7 \\
 & 4.05--4.10 & 1138 & $1.46 \pm 0.05$ & $17.56 \pm 0.48$ & 0.54 & 0.14 & 0.061 & $<0.010$ & 8.6 \\
 & 4.10--4.15 & 1238 & $1.55 \pm 0.05$ & $18.81 \pm 0.47$ & 0.54 & 0.12 & 0.040 & $<0.010$ & 8.0 \\
 & 4.15--4.20 & 1385 & $1.43 \pm 0.04$ & $20.50 \pm 0.52$ & 0.54 & 0.12 & 0.049 & $<0.010$ & 7.4 \\
 & 4.20--4.25 & 1186 & $1.64 \pm 0.05$ & $23.00 \pm 0.55$ & 0.49 & 0.09 & 0.039 & $<0.010$ & 6.4 \\
 & 4.25--4.30 & 447 & $1.52 \pm 0.09$ & $22.91 \pm 0.96$ & 0.52 & 0.10 & 0.034 & 0.320 & 6.5 \\
6450--6500 & 3.80--3.85 & 431 & $1.39 \pm 0.07$ & $31.89 \pm 1.42$ & 0.48 & 0.07 & 0.037 & 0.240 & 4.7 \\
 & 3.85--3.90 & 547 & $1.35 \pm 0.07$ & $26.59 \pm 1.10$ & 0.51 & 0.10 & 0.032 & 0.210 & 5.8 \\
 & 3.90--3.95 & 626 & $1.48 \pm 0.07$ & $24.08 \pm 0.86$ & 0.51 & 0.09 & 0.040 & 0.010 & 6.2 \\
 & 3.95--4.00 & 667 & $1.29 \pm 0.05$ & $20.41 \pm 0.82$ & 0.54 & 0.13 & 0.085 & $<0.010$ & 7.5 \\
 & 4.00--4.05 & 839 & $1.46 \pm 0.06$ & $18.68 \pm 0.60$ & 0.53 & 0.13 & 0.069 & $<0.010$ & 8.1 \\
 & 4.05--4.10 & 969 & $1.49 \pm 0.05$ & $18.92 \pm 0.55$ & 0.53 & 0.12 & 0.057 & $<0.010$ & 8.0 \\
 & 4.10--4.15 & 1233 & $1.41 \pm 0.04$ & $21.56 \pm 0.58$ & 0.52 & 0.11 & 0.055 & $<0.010$ & 7.1 \\
 & 4.15--4.20 & 1250 & $1.41 \pm 0.04$ & $24.22 \pm 0.64$ & 0.51 & 0.10 & 0.052 & $<0.010$ & 6.3 \\
 & 4.20--4.25 & 985 & $1.49 \pm 0.05$ & $27.16 \pm 0.76$ & 0.48 & 0.08 & 0.053 & $<0.010$ & 5.5 \\
 & 4.25--4.30 & 318 & $1.57 \pm 0.10$ & $24.81 \pm 1.18$ & 0.49 & 0.08 & 0.039 & 0.310 & 5.9 \\
6500--6600 & 3.70--3.80 & 670 & $1.29 \pm 0.05$ & $47.62 \pm 1.77$ & 0.44 & 0.05 & 0.047 & $<0.010$ & 3.2 \\
6500--6550 & 3.80--3.85 & 349 & $1.44 \pm 0.08$ & $38.51 \pm 1.81$ & 0.44 & 0.06 & 0.025 & 0.840 & 3.7 \\
 & 3.85--3.90 & 546 & $1.44 \pm 0.07$ & $30.55 \pm 1.17$ & 0.47 & 0.07 & 0.040 & 0.030 & 4.9 \\
 & 3.90--3.95 & 691 & $1.47 \pm 0.06$ & $27.91 \pm 0.93$ & 0.47 & 0.08 & 0.053 & $<0.010$ & 5.3 \\
 & 3.95--4.00 & 721 & $1.34 \pm 0.05$ & $22.32 \pm 0.82$ & 0.52 & 0.12 & 0.061 & $<0.010$ & 6.9 \\
 & 4.00--4.05 & 846 & $1.33 \pm 0.05$ & $19.59 \pm 0.68$ & 0.55 & 0.14 & 0.072 & $<0.010$ & 7.8 \\
 & 4.05--4.10 & 1195 & $1.51 \pm 0.05$ & $21.15 \pm 0.54$ & 0.49 & 0.11 & 0.084 & $<0.010$ & 7.1 \\
 & 4.10--4.15 & 1243 & $1.58 \pm 0.05$ & $24.09 \pm 0.57$ & 0.49 & 0.09 & 0.042 & $<0.010$ & 6.1 \\
 & 4.15--4.20 & 1241 & $1.44 \pm 0.04$ & $27.74 \pm 0.71$ & 0.48 & 0.08 & 0.047 & $<0.010$ & 5.4 \\
6500--6600 & 4.20--4.30 & 2024 & $1.48 \pm 0.03$ & $33.06 \pm 0.64$ & 0.45 & 0.06 & 0.051 & $<0.010$ & 4.4 \\
6550--6600 & 3.80--3.85 & 271 & $1.44 \pm 0.09$ & $48.10 \pm 2.52$ & 0.41 & 0.04 & 0.034 & 0.640 & 2.9 \\
 & 3.85--3.90 & 439 & $1.71 \pm 0.09$ & $34.95 \pm 1.26$ & 0.42 & 0.05 & 0.038 & 0.120 & 3.7 \\
 & 3.90--3.95 & 612 & $1.35 \pm 0.06$ & $31.88 \pm 1.22$ & 0.48 & 0.08 & 0.051 & $<0.010$ & 4.8 \\
 & 3.95--4.00 & 637 & $1.42 \pm 0.06$ & $27.79 \pm 1.00$ & 0.48 & 0.08 & 0.055 & $<0.010$ & 5.4 \\
 & 4.00--4.05 & 789 & $1.49 \pm 0.06$ & $22.49 \pm 0.72$ & 0.51 & 0.10 & 0.070 & $<0.010$ & 6.7 \\
 & 4.05--4.10 & 1091 & $1.37 \pm 0.05$ & $21.71 \pm 0.64$ & 0.53 & 0.12 & 0.069 & $<0.010$ & 7.0 \\
 & 4.10--4.15 & 1172 & $1.49 \pm 0.05$ & $25.35 \pm 0.65$ & 0.49 & 0.09 & 0.053 & $<0.010$ & 5.9 \\
 & 4.15--4.20 & 1204 & $1.43 \pm 0.04$ & $31.71 \pm 0.82$ & 0.47 & 0.07 & 0.054 & $<0.010$ & 4.7 \\
6600--6700 & 3.70--3.80 & 491 & $1.39 \pm 0.07$ & $66.06 \pm 2.63$ & 0.37 & 0.03 & 0.030 & 0.450 & 2.1 \\
 & 3.80--3.90 & 1006 & $1.41 \pm 0.05$ & $50.51 \pm 1.40$ & 0.41 & 0.04 & 0.025 & 0.130 & 2.8 \\
6600--6650 & 3.90--3.95 & 542 & $1.53 \pm 0.07$ & $37.24 \pm 1.33$ & 0.43 & 0.05 & 0.045 & 0.010 & 3.7 \\
 & 3.95--4.00 & 647 & $1.40 \pm 0.06$ & $32.66 \pm 1.18$ & 0.47 & 0.07 & 0.045 & 0.010 & 4.6 \\
 & 4.00--4.05 & 734 & $1.20 \pm 0.05$ & $27.48 \pm 1.09$ & 0.52 & 0.11 & 0.061 & $<0.010$ & 5.8 \\
 & 4.05--4.10 & 992 & $1.45 \pm 0.05$ & $25.48 \pm 0.73$ & 0.49 & 0.09 & 0.078 & $<0.010$ & 5.9 \\
 & 4.10--4.15 & 1164 & $1.39 \pm 0.04$ & $26.75 \pm 0.73$ & 0.49 & 0.09 & 0.075 & $<0.010$ & 5.7 \\
 & 4.15--4.20 & 1161 & $1.40 \pm 0.04$ & $35.64 \pm 0.95$ & 0.45 & 0.06 & 0.067 & $<0.010$ & 4.2 \\
6600--6700 & 4.20--4.30 & 1489 & $1.54 \pm 0.04$ & $41.25 \pm 0.88$ & 0.42 & 0.04 & 0.061 & $<0.010$ & 3.3 \\
6650--6700 & 3.90--3.95 & 428 & $1.51 \pm 0.08$ & $47.30 \pm 1.89$ & 0.41 & 0.04 & 0.032 & 0.450 & 2.8 \\
 & 3.95--4.00 & 520 & $1.39 \pm 0.06$ & $40.59 \pm 1.61$ & 0.44 & 0.06 & 0.045 & 0.010 & 3.6 \\
 & 4.00--4.05 & 713 & $1.44 \pm 0.06$ & $35.95 \pm 1.19$ & 0.45 & 0.06 & 0.052 & $<0.010$ & 4.0 \\
 & 4.05--4.10 & 918 & $1.27 \pm 0.04$ & $30.93 \pm 1.03$ & 0.49 & 0.09 & 0.066 & $<0.010$ & 5.1 \\
 & 4.10--4.15 & 1049 & $1.27 \pm 0.04$ & $30.51 \pm 0.95$ & 0.49 & 0.09 & 0.086 & $<0.010$ & 5.2 \\
 & 4.15--4.20 & 963 & $1.43 \pm 0.05$ & $39.18 \pm 1.12$ & 0.44 & 0.06 & 0.076 & $<0.010$ & 3.7 \\
6700--6800 & 3.70--3.80 & 528 & $1.43 \pm 0.07$ & $84.78 \pm 3.15$ & 0.28 & 0.02 & 0.032 & 0.220 & 1.4 \\
 & 3.80--3.90 & 888 & $1.61 \pm 0.06$ & $72.07 \pm 1.84$ & 0.34 & 0.02 & 0.019 & 0.710 & 1.5 \\
6700--6750 & 3.90--3.95 & 320 & $1.51 \pm 0.09$ & $56.73 \pm 2.60$ & 0.38 & 0.03 & 0.037 & 0.310 & 2.2 \\
 & 3.95--4.00 & 430 & $1.72 \pm 0.09$ & $47.42 \pm 1.68$ & 0.38 & 0.03 & 0.042 & 0.050 & 2.4 \\
 & 4.00--4.05 & 606 & $1.43 \pm 0.06$ & $44.87 \pm 1.59$ & 0.42 & 0.05 & 0.040 & 0.020 & 3.1 \\
 & 4.05--4.10 & 788 & $1.30 \pm 0.05$ & $35.98 \pm 1.25$ & 0.47 & 0.07 & 0.080 & $<0.010$ & 4.3 \\
 & 4.10--4.15 & 984 & $1.34 \pm 0.04$ & $35.22 \pm 1.06$ & 0.46 & 0.07 & 0.092 & $<0.010$ & 4.3 \\
 & 4.15--4.20 & 846 & $1.34 \pm 0.05$ & $41.45 \pm 1.33$ & 0.45 & 0.06 & 0.092 & $<0.010$ & 3.6 \\
6700--6800 & 4.20--4.30 & 922 & $1.56 \pm 0.05$ & $47.19 \pm 1.25$ & 0.40 & 0.04 & 0.059 & $<0.010$ & 2.7 \\
6750--6800 & 3.90--3.95 & 283 & $1.71 \pm 0.10$ & $65.15 \pm 2.82$ & 0.35 & 0.02 & 0.044 & 0.190 & 1.6 \\
 & 3.95--4.00 & 325 & $1.49 \pm 0.09$ & $55.09 \pm 2.54$ & 0.39 & 0.03 & 0.042 & 0.150 & 2.3 \\
 & 4.00--4.05 & 450 & $1.48 \pm 0.07$ & $51.13 \pm 2.03$ & 0.40 & 0.04 & 0.031 & 0.450 & 2.6 \\
 & 4.05--4.10 & 709 & $1.49 \pm 0.06$ & $45.28 \pm 1.43$ & 0.41 & 0.04 & 0.048 & $<0.010$ & 3.0 \\
 & 4.10--4.15 & 931 & $1.34 \pm 0.05$ & $44.39 \pm 1.36$ & 0.44 & 0.05 & 0.061 & $<0.010$ & 3.4 \\
 & 4.15--4.20 & 821 & $1.39 \pm 0.05$ & $44.14 \pm 1.39$ & 0.43 & 0.05 & 0.077 & $<0.010$ & 3.3 \\
6800--6900 & 3.70--3.80 & 491 & $1.48 \pm 0.08$ & $96.74 \pm 3.65$ & 0.20 & 0.02 & 0.030 & 0.430 & 1.2 \\
 & 3.80--3.90 & 861 & $1.68 \pm 0.06$ & $92.13 \pm 2.30$ & 0.28 & 0.01 & 0.028 & 0.170 & 1.0 \\
6800--6850 & 3.90--3.95 & 326 & $1.88 \pm 0.11$ & $77.87 \pm 2.87$ & 0.32 & 0.01 & 0.032 & 0.640 & 1.0 \\
 & 3.95--4.00 & 320 & $1.54 \pm 0.09$ & $62.93 \pm 2.82$ & 0.37 & 0.03 & 0.043 & 0.100 & 1.9 \\
 & 4.00--4.05 & 398 & $1.54 \pm 0.08$ & $52.09 \pm 2.11$ & 0.39 & 0.03 & 0.042 & 0.050 & 2.4 \\
 & 4.05--4.10 & 532 & $1.59 \pm 0.07$ & $54.42 \pm 1.85$ & 0.38 & 0.03 & 0.028 & 0.490 & 2.2 \\
 & 4.10--4.15 & 738 & $1.42 \pm 0.05$ & $56.52 \pm 1.80$ & 0.40 & 0.03 & 0.060 & $<0.010$ & 2.4 \\
 & 4.15--4.20 & 751 & $1.47 \pm 0.05$ & $53.56 \pm 1.65$ & 0.40 & 0.03 & 0.070 & $<0.010$ & 2.5 \\
6800--6900 & 4.20--4.30 & 722 & $1.45 \pm 0.06$ & $56.53 \pm 1.79$ & 0.39 & 0.03 & 0.055 & $<0.010$ & 2.3 \\
6850--6900 & 3.90--3.95 & 277 & $1.65 \pm 0.11$ & $78.34 \pm 3.51$ & 0.32 & 0.02 & 0.061 & $<0.010$ & 1.3 \\
 & 3.95--4.00 & 275 & $1.72 \pm 0.11$ & $72.13 \pm 3.14$ & 0.34 & 0.02 & 0.036 & 0.510 & 1.3 \\
 & 4.00--4.05 & 354 & $1.56 \pm 0.09$ & $66.91 \pm 2.81$ & 0.36 & 0.02 & 0.034 & 0.370 & 1.7 \\
 & 4.05--4.10 & 458 & $1.60 \pm 0.08$ & $66.43 \pm 2.39$ & 0.36 & 0.02 & 0.033 & 0.270 & 1.7 \\
 & 4.10--4.15 & 518 & $1.56 \pm 0.07$ & $66.76 \pm 2.32$ & 0.36 & 0.02 & 0.024 & 0.750 & 1.7 \\
 & 4.15--4.20 & 521 & $1.63 \pm 0.08$ & $69.31 \pm 2.30$ & 0.35 & 0.02 & 0.027 & 0.570 & 1.5 \\
6900--7000 & 3.70--3.80 & 446 & $1.74 \pm 0.10$ & $115.48 \pm 3.98$ & 0.13 & 0.01 & 0.051 & $<0.010$ & 0.6 \\
 & 3.80--3.90 & 782 & $1.68 \pm 0.07$ & $103.38 \pm 2.73$ & 0.21 & 0.01 & 0.029 & 0.140 & 0.8 \\
6900--6950 & 3.90--3.95 & 337 & $1.69 \pm 0.10$ & $88.00 \pm 3.48$ & 0.30 & 0.01 & 0.034 & 0.500 & 1.0 \\
 & 3.95--4.00 & 302 & $1.91 \pm 0.12$ & $85.28 \pm 3.22$ & 0.31 & 0.01 & 0.032 & 0.680 & 0.9 \\
 & 4.00--4.05 & 366 & $1.52 \pm 0.09$ & $68.54 \pm 2.90$ & 0.36 & 0.02 & 0.034 & 0.380 & 1.7 \\
 & 4.05--4.10 & 439 & $1.56 \pm 0.08$ & $73.86 \pm 2.78$ & 0.34 & 0.02 & 0.048 & 0.010 & 1.5 \\
 & 4.10--4.15 & 490 & $1.75 \pm 0.08$ & $80.18 \pm 2.56$ & 0.32 & 0.01 & 0.023 & 0.820 & 1.1 \\
 & 4.15--4.20 & 408 & $1.75 \pm 0.09$ & $76.97 \pm 2.70$ & 0.33 & 0.01 & 0.045 & 0.030 & 1.2 \\
6900--7000 & 4.20--4.30 & 455 & $1.88 \pm 0.09$ & $72.12 \pm 2.25$ & 0.33 & 0.01 & 0.032 & 0.360 & 1.1 \\
6950--7000 & 3.90--3.95 & 297 & $2.01 \pm 0.13$ & $97.89 \pm 3.56$ & 0.29 & 0.01 & 0.037 & 0.500 & 0.6 \\
 & 3.95--4.00 & 312 & $1.84 \pm 0.12$ & $90.94 \pm 3.47$ & 0.30 & 0.01 & 0.054 & $<0.010$ & 0.8 \\
 & 4.00--4.05 & 375 & $1.67 \pm 0.09$ & $79.10 \pm 3.00$ & 0.32 & 0.01 & 0.022 & 0.840 & 1.2 \\
 & 4.05--4.10 & 404 & $1.66 \pm 0.09$ & $74.67 \pm 2.76$ & 0.34 & 0.02 & 0.038 & 0.180 & 1.3 \\
 & 4.10--4.15 & 408 & $1.79 \pm 0.09$ & $84.63 \pm 2.91$ & 0.32 & 0.01 & 0.028 & 0.630 & 1.0 \\
 & 4.15--4.20 & 330 & $1.75 \pm 0.11$ & $91.47 \pm 3.56$ & 0.29 & 0.01 & 0.052 & 0.070 & 0.9 \\
7000--7100 & 3.60--3.70 & 276 & $1.59 \pm 0.13$ & $134.00 \pm 7.20$ & $-0.16$ & 0.03 & 0.064 & 0.010 & 0.6 \\
 & 3.70--3.80 & 533 & $1.68 \pm 0.10$ & $133.47 \pm 4.72$ & $-0.09$ & 0.02 & 0.050 & $<0.010$ & 0.6 \\
 & 3.80--3.90 & 967 & $1.79 \pm 0.07$ & $118.55 \pm 2.71$ & 0.14 & 0.01 & 0.030 & 0.100 & 0.6 \\
7000--7050 & 3.90--3.95 & 354 & $2.18 \pm 0.13$ & $113.94 \pm 3.55$ & 0.27 & 0.00 & 0.033 & 0.420 & 0.4 \\
 & 3.95--4.00 & 375 & $2.10 \pm 0.12$ & $101.92 \pm 3.18$ & 0.29 & 0.00 & 0.046 & 0.030 & 0.5 \\
 & 4.00--4.05 & 384 & $1.76 \pm 0.10$ & $89.38 \pm 3.20$ & 0.30 & 0.01 & 0.033 & 0.440 & 0.9 \\
 & 4.05--4.10 & 529 & $1.79 \pm 0.08$ & $79.63 \pm 2.40$ & 0.32 & 0.01 & 0.018 & 0.960 & 1.1 \\
 & 4.10--4.15 & 490 & $1.83 \pm 0.09$ & $85.02 \pm 2.61$ & 0.31 & 0.01 & 0.029 & 0.460 & 0.9 \\
 & 4.15--4.20 & 408 & $2.03 \pm 0.11$ & $96.09 \pm 2.95$ & 0.30 & 0.01 & 0.026 & 0.830 & 0.6 \\
7000--7100 & 4.20--4.30 & 402 & $1.95 \pm 0.11$ & $90.65 \pm 2.91$ & 0.30 & 0.01 & 0.024 & 0.840 & 0.7 \\
7050--7100 & 3.90--3.95 & 348 & $1.98 \pm 0.12$ & $114.10 \pm 3.90$ & 0.23 & 0.01 & 0.034 & 0.390 & 0.5 \\
 & 3.95--4.00 & 380 & $1.91 \pm 0.11$ & $110.23 \pm 3.72$ & 0.24 & 0.01 & 0.026 & 0.710 & 0.6 \\
 & 4.00--4.05 & 475 & $1.84 \pm 0.09$ & $98.72 \pm 3.07$ & 0.28 & 0.01 & 0.024 & 0.780 & 0.7 \\
 & 4.05--4.10 & 600 & $1.72 \pm 0.08$ & $88.66 \pm 2.60$ & 0.30 & 0.01 & 0.024 & 0.570 & 1.0 \\
 & 4.10--4.15 & 515 & $1.65 \pm 0.08$ & $93.30 \pm 3.06$ & 0.27 & 0.01 & 0.028 & 0.530 & 1.0 \\
 & 4.15--4.20 & 429 & $1.75 \pm 0.09$ & $94.52 \pm 3.22$ & 0.28 & 0.01 & 0.024 & 0.830 & 0.9 \\
7100--7200 & 3.60--3.70 & 275 & $1.99 \pm 0.17$ & $160.14 \pm 7.42$ & $-0.22$ & 0.03 & 0.042 & 0.240 & 0.3 \\
 & 3.70--3.80 & 547 & $1.63 \pm 0.11$ & $162.90 \pm 7.70$ & $-0.43$ & 0.05 & 0.048 & 0.010 & 0.5 \\
 & 3.80--3.90 & 996 & $1.65 \pm 0.07$ & $135.86 \pm 3.67$ & $-0.13$ & 0.02 & 0.031 & 0.040 & 0.6 \\
7100--7150 & 3.90--3.95 & 334 & $1.94 \pm 0.12$ & $121.88 \pm 4.40$ & 0.17 & 0.01 & 0.033 & 0.540 & 0.5 \\
 & 3.95--4.00 & 358 & $1.95 \pm 0.12$ & $117.94 \pm 4.05$ & 0.20 & 0.01 & 0.030 & 0.500 & 0.5 \\
 & 4.00--4.05 & 433 & $2.11 \pm 0.12$ & $112.45 \pm 3.26$ & 0.26 & 0.00 & 0.031 & 0.430 & 0.4 \\
 & 4.05--4.10 & 571 & $1.64 \pm 0.08$ & $95.69 \pm 3.02$ & 0.25 & 0.01 & 0.024 & 0.720 & 1.0 \\
 & 4.10--4.15 & 489 & $1.65 \pm 0.08$ & $95.12 \pm 3.21$ & 0.26 & 0.01 & 0.030 & 0.360 & 1.0 \\
 & 4.15--4.20 & 413 & $1.75 \pm 0.09$ & $94.28 \pm 3.29$ & 0.29 & 0.01 & 0.036 & 0.250 & 0.9 \\
7100--7200 & 4.20--4.30 & 471 & $1.66 \pm 0.08$ & $85.54 \pm 2.91$ & 0.30 & 0.01 & 0.041 & 0.080 & 1.1 \\
7150--7200 & 3.90--3.95 & 365 & $2.08 \pm 0.13$ & $127.37 \pm 4.13$ & 0.17 & 0.01 & 0.045 & 0.060 & 0.4 \\
 & 3.95--4.00 & 367 & $1.91 \pm 0.12$ & $125.27 \pm 4.41$ & 0.14 & 0.01 & 0.031 & 0.590 & 0.5 \\
 & 4.00--4.05 & 459 & $2.08 \pm 0.11$ & $118.66 \pm 3.39$ & 0.23 & 0.00 & 0.024 & 0.790 & 0.4 \\
 & 4.05--4.10 & 567 & $1.90 \pm 0.09$ & $114.67 \pm 3.20$ & 0.21 & 0.01 & 0.026 & 0.670 & 0.5 \\
 & 4.10--4.15 & 568 & $1.79 \pm 0.08$ & $106.50 \pm 3.11$ & 0.23 & 0.01 & 0.027 & 0.440 & 0.7 \\
 & 4.15--4.20 & 444 & $1.77 \pm 0.10$ & $98.31 \pm 3.27$ & 0.26 & 0.01 & 0.042 & 0.060 & 0.8 \\
7200--7300 & 3.70--3.80 & 400 & $2.10 \pm 0.16$ & $166.69 \pm 6.24$ & $-0.25$ & 0.03 & 0.049 & 0.050 & 0.2 \\
 & 3.80--3.90 & 821 & $2.14 \pm 0.10$ & $160.85 \pm 3.86$ & $-0.13$ & 0.02 & 0.024 & 0.330 & 0.2 \\
7200--7250 & 3.90--3.95 & 298 & $2.59 \pm 0.20$ & $151.63 \pm 4.62$ & 0.16 & 0.00 & 0.026 & 0.900 & 0.2 \\
 & 3.95--4.00 & 294 & $1.58 \pm 0.12$ & $136.38 \pm 7.22$ & $-0.16$ & 0.03 & 0.050 & 0.010 & 0.6 \\
 & 4.00--4.05 & 459 & $1.86 \pm 0.10$ & $122.84 \pm 3.96$ & 0.13 & 0.01 & 0.034 & 0.260 & 0.5 \\
 & 4.05--4.10 & 487 & $1.97 \pm 0.10$ & $117.51 \pm 3.42$ & 0.21 & 0.01 & 0.032 & 0.370 & 0.5 \\
 & 4.10--4.15 & 543 & $1.78 \pm 0.08$ & $107.53 \pm 3.24$ & 0.22 & 0.01 & 0.022 & 0.840 & 0.7 \\
 & 4.15--4.20 & 429 & $1.78 \pm 0.10$ & $108.39 \pm 3.67$ & 0.22 & 0.01 & 0.026 & 0.760 & 0.7 \\
7200--7300 & 4.20--4.30 & 426 & $1.92 \pm 0.10$ & $92.73 \pm 2.93$ & 0.30 & 0.01 & 0.042 & 0.080 & 0.7 \\
7250--7300 & 3.90--3.95 & 288 & $2.03 \pm 0.15$ & $148.93 \pm 6.04$ & $-0.04$ & 0.02 & 0.037 & 0.440 & 0.3 \\
 & 3.95--4.00 & 277 & $1.81 \pm 0.14$ & $138.56 \pm 6.31$ & $-0.06$ & 0.02 & 0.051 & 0.050 & 0.4 \\
 & 4.00--4.05 & 493 & $2.13 \pm 0.12$ & $132.03 \pm 3.63$ & 0.16 & 0.01 & 0.036 & 0.120 & 0.3 \\
 & 4.05--4.10 & 516 & $1.83 \pm 0.10$ & $120.32 \pm 3.69$ & 0.14 & 0.01 & 0.031 & 0.300 & 0.5 \\
 & 4.10--4.15 & 503 & $1.91 \pm 0.10$ & $128.13 \pm 3.89$ & 0.11 & 0.01 & 0.042 & 0.030 & 0.4 \\
 & 4.15--4.20 & 440 & $1.86 \pm 0.10$ & $115.78 \pm 3.75$ & 0.19 & 0.01 & 0.030 & 0.500 & 0.6 \\
7300--7400 & 3.70--3.80 & 408 & $1.69 \pm 0.14$ & $194.22 \pm 13.76$ & $-0.61$ & 0.09 & 0.053 & $<0.010$ & 0.3 \\
 & 3.80--3.90 & 760 & $1.88 \pm 0.11$ & $183.40 \pm 7.04$ & $-0.49$ & 0.06 & 0.055 & $<0.010$ & 0.3 \\
 & 3.90--4.00 & 1058 & $1.95 \pm 0.08$ & $146.18 \pm 3.21$ & $-0.06$ & 0.02 & 0.038 & $<0.010$ & 0.3 \\
7300--7350 & 4.00--4.05 & 355 & $1.92 \pm 0.13$ & $135.71 \pm 5.00$ & 0.03 & 0.01 & 0.028 & 0.770 & 0.4 \\
 & 4.05--4.10 & 435 & $1.95 \pm 0.11$ & $134.01 \pm 4.34$ & 0.08 & 0.01 & 0.031 & 0.360 & 0.4 \\
 & 4.10--4.15 & 531 & $1.85 \pm 0.10$ & $130.82 \pm 4.04$ & 0.06 & 0.01 & 0.029 & 0.380 & 0.5 \\
 & 4.15--4.20 & 413 & $1.71 \pm 0.10$ & $114.68 \pm 4.17$ & 0.14 & 0.01 & 0.042 & 0.030 & 0.7 \\
7300--7400 & 4.20--4.30 & 411 & $1.78 \pm 0.10$ & $101.75 \pm 3.51$ & 0.25 & 0.01 & 0.029 & 0.540 & 0.7 \\
7350--7400 & 4.00--4.05 & 323 & $1.70 \pm 0.13$ & $137.92 \pm 6.36$ & $-0.13$ & 0.02 & 0.040 & 0.240 & 0.5 \\
 & 4.05--4.10 & 413 & $2.02 \pm 0.12$ & $129.67 \pm 4.10$ & 0.14 & 0.01 & 0.037 & 0.150 & 0.4 \\
 & 4.10--4.15 & 350 & $1.81 \pm 0.12$ & $136.90 \pm 5.50$ & $-0.03$ & 0.02 & 0.023 & 0.920 & 0.5 \\
 & 4.15--4.20 & 312 & $1.85 \pm 0.12$ & $119.66 \pm 4.66$ & 0.16 & 0.01 & 0.029 & 0.780 & 0.5 \\
7400--7500 & 3.70--3.80 & 424 & $1.99 \pm 0.17$ & $191.37 \pm 10.04$ & $-0.54$ & 0.06 & 0.040 & 0.130 & 0.2 \\
 & 3.80--3.90 & 809 & $2.10 \pm 0.11$ & $171.48 \pm 4.68$ & $-0.28$ & 0.03 & 0.046 & $<0.010$ & 0.2 \\
 & 3.90--4.00 & 1008 & $1.99 \pm 0.09$ & $157.58 \pm 3.72$ & $-0.17$ & 0.02 & 0.032 & 0.010 & 0.3 \\
7400--7450 & 4.00--4.05 & 279 & $1.62 \pm 0.14$ & $150.14 \pm 8.91$ & $-0.32$ & 0.04 & 0.051 & 0.110 & 0.5 \\
 & 4.05--4.10 & 371 & $2.03 \pm 0.13$ & $137.50 \pm 4.67$ & 0.08 & 0.01 & 0.042 & 0.130 & 0.3 \\
 & 4.10--4.15 & 353 & $1.97 \pm 0.12$ & $134.14 \pm 4.77$ & 0.09 & 0.01 & 0.039 & 0.190 & 0.4 \\
 & 4.15--4.20 & 286 & $1.71 \pm 0.13$ & $125.99 \pm 5.73$ & 0.02 & 0.02 & 0.060 & 0.020 & 0.6 \\
7400--7500 & 4.20--4.30 & 405 & $1.93 \pm 0.11$ & $114.18 \pm 3.72$ & 0.22 & 0.01 & 0.035 & 0.270 & 0.5 \\
7450--7500 & 4.00--4.05 & 332 & $2.30 \pm 0.18$ & $142.92 \pm 4.54$ & 0.12 & 0.01 & 0.079 & $<0.010$ & 0.2 \\
 & 4.05--4.10 & 426 & $2.06 \pm 0.13$ & $145.89 \pm 4.69$ & $-0.01$ & 0.01 & 0.037 & 0.200 & 0.3 \\
 & 4.10--4.15 & 441 & $1.75 \pm 0.11$ & $140.23 \pm 5.36$ & $-0.09$ & 0.02 & 0.030 & 0.460 & 0.5 \\
 & 4.15--4.20 & 384 & $1.90 \pm 0.11$ & $122.70 \pm 4.21$ & 0.16 & 0.01 & 0.037 & 0.220 & 0.5 \\
7500--7600 & 3.70--3.80 & 341 & $2.01 \pm 0.16$ & $175.51 \pm 8.22$ & $-0.36$ & 0.04 & 0.041 & 0.160 & 0.2 \\
 & 3.80--3.90 & 604 & $1.88 \pm 0.12$ & $175.87 \pm 6.97$ & $-0.41$ & 0.05 & 0.043 & $<0.010$ & 0.3 \\
 & 3.90--4.00 & 746 & $1.97 \pm 0.10$ & $162.31 \pm 4.72$ & $-0.25$ & 0.03 & 0.048 & $<0.010$ & 0.3 \\
 & 4.00--4.10 & 1027 & $2.02 \pm 0.08$ & $147.11 \pm 3.15$ & $-0.04$ & 0.01 & 0.044 & $<0.010$ & 0.3 \\
 & 4.10--4.20 & 1041 & $1.94 \pm 0.08$ & $140.38 \pm 3.03$ & 0.00 & 0.01 & 0.031 & 0.010 & 0.4 \\
 & 4.20--4.30 & 335 & $1.87 \pm 0.11$ & $102.50 \pm 3.74$ & 0.27 & 0.01 & 0.038 & 0.330 & 0.7 \\
7600--7700 & 3.70--3.80 & 257 & $1.60 \pm 0.16$ & $171.36 \pm 13.16$ & $-0.50$ & 0.07 & 0.056 & 0.030 & 0.5 \\
 & 3.80--3.90 & 537 & $2.05 \pm 0.13$ & $168.93 \pm 5.74$ & $-0.28$ & 0.03 & 0.035 & 0.080 & 0.2 \\
 & 3.90--4.00 & 629 & $2.03 \pm 0.11$ & $162.05 \pm 4.85$ & $-0.19$ & 0.02 & 0.038 & 0.050 & 0.3 \\
 & 4.00--4.10 & 915 & $1.94 \pm 0.08$ & $145.31 \pm 3.44$ & $-0.06$ & 0.02 & 0.047 & $<0.010$ & 0.3 \\
 & 4.10--4.20 & 952 & $1.88 \pm 0.08$ & $143.01 \pm 3.43$ & $-0.05$ & 0.02 & 0.041 & $<0.010$ & 0.4 \\
 & 4.20--4.30 & 376 & $1.95 \pm 0.11$ & $115.36 \pm 3.85$ & 0.22 & 0.01 & 0.051 & 0.020 & 0.5 \\
7700--7800 & 3.80--3.90 & 481 & $1.92 \pm 0.15$ & $187.63 \pm 9.30$ & $-0.53$ & 0.06 & 0.051 & $<0.010$ & 0.2 \\
 & 3.90--4.00 & 584 & $1.94 \pm 0.12$ & $164.07 \pm 5.60$ & $-0.27$ & 0.03 & 0.046 & 0.010 & 0.3 \\
 & 4.00--4.10 & 812 & $1.91 \pm 0.10$ & $170.12 \pm 5.31$ & $-0.33$ & 0.04 & 0.036 & $<0.010$ & 0.3 \\
 & 4.10--4.20 & 789 & $1.92 \pm 0.09$ & $148.53 \pm 3.93$ & $-0.10$ & 0.02 & 0.033 & 0.010 & 0.3 \\
 & 4.20--4.30 & 355 & $1.88 \pm 0.11$ & $105.85 \pm 3.74$ & 0.25 & 0.01 & 0.064 & $<0.010$ & 0.6 \\
7800--7900 & 3.80--3.90 & 461 & $2.10 \pm 0.15$ & $174.22 \pm 6.49$ & $-0.33$ & 0.03 & 0.061 & $<0.010$ & 0.2 \\
 & 3.90--4.00 & 614 & $2.06 \pm 0.11$ & $162.98 \pm 4.86$ & $-0.18$ & 0.02 & 0.037 & 0.030 & 0.2 \\
 & 4.00--4.10 & 816 & $2.01 \pm 0.10$ & $164.53 \pm 4.52$ & $-0.25$ & 0.03 & 0.041 & $<0.010$ & 0.3 \\
 & 4.10--4.20 & 757 & $1.97 \pm 0.09$ & $152.13 \pm 4.04$ & $-0.11$ & 0.02 & 0.034 & 0.040 & 0.3 \\
 & 4.20--4.30 & 385 & $1.97 \pm 0.11$ & $118.22 \pm 3.89$ & 0.21 & 0.01 & 0.051 & 0.010 & 0.5 \\
7900--8000 & 3.80--3.90 & 436 & $1.83 \pm 0.13$ & $169.03 \pm 7.64$ & $-0.36$ & 0.04 & 0.038 & 0.040 & 0.3 \\
 & 3.90--4.00 & 642 & $1.72 \pm 0.10$ & $167.66 \pm 6.84$ & $-0.40$ & 0.05 & 0.034 & 0.050 & 0.4 \\
 & 4.00--4.10 & 810 & $2.13 \pm 0.10$ & $161.57 \pm 3.95$ & $-0.14$ & 0.02 & 0.029 & 0.170 & 0.2 \\
 & 4.10--4.20 & 784 & $2.06 \pm 0.09$ & $147.35 \pm 3.52$ & $-0.01$ & 0.01 & 0.045 & $<0.010$ & 0.3 \\
 & 4.20--4.30 & 449 & $2.02 \pm 0.10$ & $114.04 \pm 3.37$ & 0.24 & 0.01 & 0.029 & 0.480 & 0.5 \\
 
\end{longtable}
\tablefoot{Sorted by \Teff, then \logg. A blank bin repeats the value above. Never-subdivided parent cells appear once with their $100~\mathrm{K} \times 0.1~\dex$ bounds. Column $p$ is the fraction of a cell's 100 artificial cells with a $D_{\rm KS}$ at least as large as that of the real cell (Sect.~\ref{sec:grid-fitting}), and $<0.010$ means that none of them was. Column $p_{78}$ is the share of the fitted distribution that falls in the $7$--$8$~\kms{} band of the pile-up (Sect.~\ref{sec:instrument-pipeline}), as a percentage of the mass inside the $5$--$275$~\kms{} window, $p_{78}=P(7\le\mathtt{vbroad}\le 8)/P(5\le\mathtt{vbroad}\le 275)$. It is a property of the model, not the share of the data in the band.}
}

\end{appendix}

\end{document}